\documentclass[]{spie}  

\usepackage[perpage]{footmisc}
\usepackage{subcaption} 
\usepackage{amsmath,amsfonts,amssymb}
\usepackage{graphicx}
\usepackage[colorlinks=true, allcolors=blue]{hyperref}
\usepackage[numbers]{natbib}

\title{Tiny-box: A tool for the versatile development and characterization of low noise fast X-ray imaging detectors}

\author[a]{Tanmoy Chattopadhyay}
\author[a]{Sven Herrmann}
\author[a,b,c]{Steven Allen}
\author[a,d]{Jack Hirschman}
\author[a,c]{Glenn Morris}
\author[e]{Marshall Bautz}
\author[e]{Andrew Malonis}
\author[e]{Richard Foster}
\author[e]{Gregory Prigozhin}
\author[f]{Dave Craig}
\author[f]{Barry Burke}

\affil[a]{Kavli Institute of Astrophysics and Cosmology, Stanford University, 452 Lomita Mall, Stanford, CA 94305, USA}
\affil[b]{Department of Physics, Stanford University, 382 Via Pueblo Mall, Stanford CA 94305, USA}
\affil[c]{SLAC National Accelerator Laboratory, 2575 Sand Hill Road, Menlo Park, CA 94025, USA}
\affil[d]{Department of Applied Physics, Stanford University, 348 Via Pueblo, Stanford, CA 94305, USA}
\affil[e]{Kavli Institute for Astrophysics and Space Research, Massachusetts Institute of Technology, Cambridge, MA USA}
\affil[f]{MIT Lincoln Laboratory, Lexington, MA, USA}

\authorinfo{Further author information: (Send correspondence to T. Chattopadhyay)\\T.C.: E-mail: tanmoyc@stanford.edu, Telephone: 1 814 852 9733}

\begin{document} 
\maketitle

\begin{abstract}
X-ray Charge Coupled Devices (CCDs) have been the workhorse for soft X-ray astronomical instruments for the past quarter century. They provide broad energy response, extremely low electronic read noise, and good energy resolution in soft X-rays. These properties, along with the large arrays and small pixel sizes available with modern-day CCDs, make them a potential candidate for next generation astronomical X-ray missions equipped with large collecting areas, high angular resolutions and wide fields of view, enabling observation of the faint, diffuse and high redshift X-ray universe. However, such high collecting area (about 30 times Chandra) requires these detectors to have an order of magnitude faster readout than current CCDs to avoid saturation and pile up effects. In this context, Stanford University and MIT have initiated the development of fast readout X-ray cameras. As a tool for this development, we have designed a fast readout, low noise electronics board (intended to work at a 5 Megapixel per second data rate) coupled with an STA Archon controller to readout a 512 $\times$ 512 CCD (from MIT Lincoln Laboratory). This versatile setup allows us to study a number of parameters and operation conditions including the option for digital shaping. In this paper, we describe the characterization test stand, the concept and development of the readout electronics, and simulation results. We also report the first measurements of read noise, energy resolution and other parameters from this set up. While this is very much a prototype, we plan to use larger, multi-node CCD devices in the future with dedicated ASIC readout systems to enable faster, parallel readout of the CCDs.
\end{abstract}


\section{INTRODUCTION}
\label{sec:intro}  
X-ray Charge Coupled Devices (CCDs) \citep{Lesser15_ccd,gruner02_ccd} 
have been the workhorse for soft X-ray instrumentation
for more than two decades. 
The excellent energy resolution, very low electronic
read noise, small pixel sizes and broad energy response of up to tens of keV of X-ray CCDs make them an obvious choice for developing sensitive soft X-ray spectro-imagers. 
The existing and past astronomical missions (e.g., ASCA, Chandra, XMM-Newton, Swift, Integral, Hitomi, and AstroSat) utilizing CCD based spectro-imagers have been extremely 
successful, and in particular Chandra, with high angular resolution of the mirrors, has been able to utilize the low pixel sizes of CCDs to produce high resolution X-ray images of astronomical sources. 

Next generation X-ray astronomical missions will aim to extend the success of Chandra by combining fine angular resolution with large collecting area to probe deeper into the high redshift and low luminosity X-ray universe \citep{vikhlinin12_smartx}. For example, the 
Lynx Observatory \citep{gaskin15_lynx}, proposed to NASA's 2020 Decadal Survey, plans to have $\sim$30 times larger collecting area than Chandra. 
Such missions will require large, focal plane X-ray detectors with an order of magnitude faster  
readout than existing X-ray CCDs to overcome pile up and saturation effects \citep{lumb00_pileup_xmm}. 

The past decade has seen significant progress in the development of fast readout, small pixel size, low power X-ray detectors
(see \cite{falcone18_HXDI} for an overview).
Active Pixel Sensors (APS) such as X-ray Hybrid CMOS detectors \citep{bai08,hull17,chattopadhyay18_HCDoverview,hull18_small_pixel}, the depleted field effect transistor (DEPFET) sensors being developed for the ATHENA Wide-Field Imager (WFI, \citep{norbert16_wfi}), and Silicon-On-Insulator technology \citep{SOI18} promise to fulfill some of these requirements. 
While most of the requirements are also met with modern X-ray CCDs, major challenges to increase the readout speed and operate at a lower power remain. 
Recently, \cite{bautz18,bautz19} have reported interesting progress in the development of low power, high speed X-ray CCDs, namely digital CCDs. 
MIT Lincoln laboratory (MITL) is developing technology to produce CCDs with 2-stage on-chip amplifier (P-JFET-based first-stage source follower with a second-stage N-MOSFET buffer for added bandwidth) to support high readout speed. On the other hand, CCD gate electrodes, formed with single-level polysilicon process, enable low power operation of the CCDs. 
These new generation chips also provide current readout mode through a newly developed Single-Electron Sensitive Readout (SiSeRO) amplifier (similar to that of DEPFET architecture), intended to provide higher responsivity and significantly better noise performance (sub-electron read noise) even at readout speeds $>$2 MHz. An advantage of the current readout mode is that the readout speed of the system is not limited by the RC time constant (1/g$_m$, where g$_m$ is the transconductance) of the device and  the total input capacitance of the readout electronics (C).     
Stanford University is 
collaborating with MIT and MIT Lincoln Laboratory to develop
fast readout electronic boards to both run these CCDs and optimize their noise performance at high readout speeds. While the details of the CCDs have been discussed in \citep{bautz18} and \cite{bautz19}, here we describe the newly developed experiment test stand designed to run the CCDs at high readout speeds and characterize their spectroscopic and noise performance.
The test stand is small (hence ``tiny-box"), around 10 cm in size in the x, y, z-directions. Successful implementation of high-speed readout of the CCDs depends on various factors including large bandwidth and low input capacitance of the amplifiers, low parasitic capacitance in the readout board, as well as the high g$_m$ of the on-chip amplifier in the CCDs. System noise also increases with the increase in bandwidth ($\sqrt{BW}$). All these factors impose serious restrictions on the readout circuitry, the choice of amplifiers and their optimization for overall high-bandwidth, low-noise performance.     
We discuss the considerations behind the development of our readout circuitry and electronic boards to support simultaneous high readout speed and low readout noise.  

Sec. \ref{stand} describes the characterization test stand and Sec. \ref{readout} the readout electronic module. For our initial studies, We have characterized an MITL DCCD (CCID85B series) at 2 MHz readout speed. The measurements use the P-JFET on-chip amplifier output. Energy resolution and read noise measurements are presented in Sec. \ref{results}. 
The results are promising and, in near future, we plan to run the CCDs at even higher readout speeds (minimum 5 MHz) to establish the high-speed readout circuitry and characterize the noise performance. 
Another goal of the project is to develop a readout module for current (drain) readout of the SiSeRO amplifiers,  
which are anticipated to provide an order of magnitude improvement in performance over the P-JFET based CCDs.  
We also plan to develop ASIC-based highly parallel signal processing chains to enable the high frame rates required by next generation X-ray astronomical missions. 

\section{Characterization test stand}
\begin{figure}
    \centering
    \includegraphics[width=0.9\linewidth]{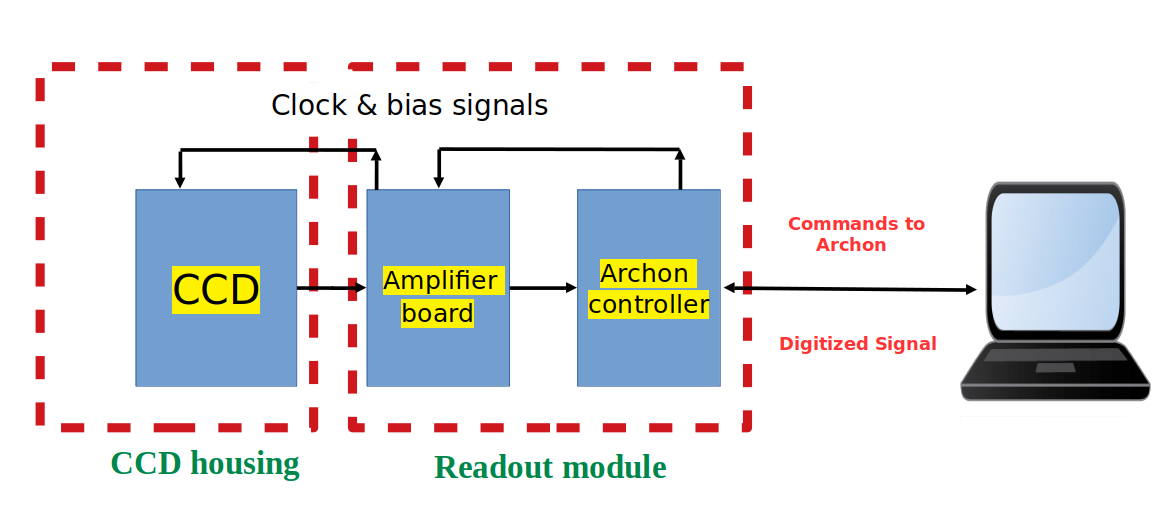}
    \caption{Schematic block diagram of the characterization test stand. The major three components of the set up are shown $-$ the CCD housing, Preamplifier board and Archon controller. The last two components make the readout module, to run the CCDs and read the output signals.}
    \label{block}
\end{figure}

In this section, we discuss the details of the experiment setup, the characterization test stand, along with the readout module developed for X-ray CCDs. Fig. \ref{block} shows the basic experiment block, the different components of the test stand and the overall concept of how to read the CCDs and characterize them.
The characterization test stand consists of two separate components: the CCD housing and the readout module. 
The CCD vacuum chamber houses the X-ray CCDs and provides cooling of the CCDs to low temperatures. An X-ray entrance window allows X-ray photons to illuminate the CCDs.
The readout module consists of two components $-$ a fast readout and low noise preamplifier board and an Archon CCD controller \citep{archon14}. 
The Archon controller provides the required bias and clock signals to run the X-ray CCDs through the preamplifier board. The preamplifier amplifies and drives the CCD analog output signal to the differential ADCs in the Archon controller. The Archon is programmed to extract the source signal amplitude from the digitally sampled video waveform by taking the difference between the signal and baseline of the waveform, which is then used to generate the images and spectra of the X-ray photons. We discuss each of these components in more detail below.

\subsection{CCD housing}
\label{stand}
\begin{figure}
    \centering
    \includegraphics[width=0.48\linewidth]{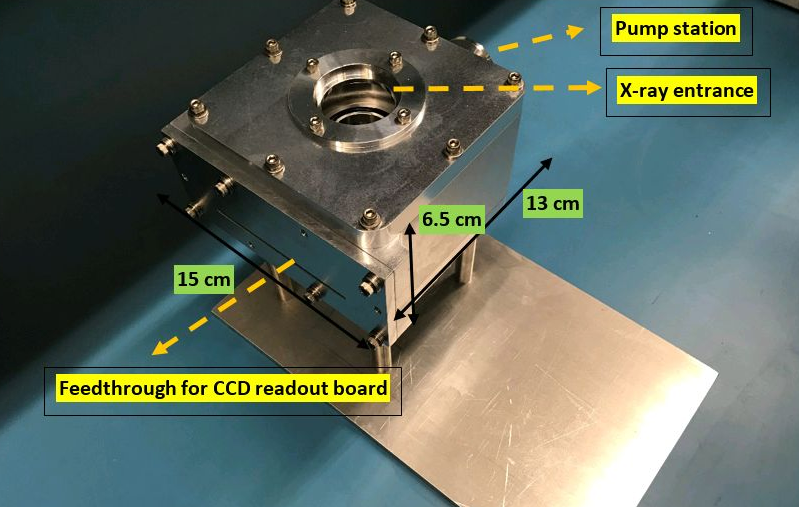}
    \caption{CCD housing: we use a small compact chamber, made of aluminum. One of the side flanges provides feedthrough for the CCD preamplfier board. A beryllium X-ray entrance window is mounted on the top flange.}
    \label{chamber}
\end{figure}

The CCD housing, shown in Fig. \ref{chamber}, is a compact vacuum chamber (13 cm $\times$ 15 cm $\times$ 6.5 cm) made of aluminum. 
The small sizes of the CCDs under test allow the chamber to be made compact. 
The X-ray detector is mounted in the middle of the chamber on an aluminum cold block (or cold finger) and faces upward towards the top flange. 
As shown in Fig. \ref{chamber}, the side flange has a narrow slot for the CCD readout board (preamplifier board) to be epoxied such that nearly half of the board with CCD mount is inside the chamber and other half is outside the chamber and connected to the next chain of the readout module (Archon controller). 
The advantage of such a compact chamber size and readout board geometry is the flexibility of using it in multiple applications, for example in an X-ray beam line for detailed spectral characterization of the CCDs, as an imager at the focal plane of an X-ray mirror, or in a proton beam line for radiation hardness tests.
The flange opposite to the readout board flange is connected to the pump station through a custom-made adapter. 
The bottom flange has a feed-through for a DB9 connector to accommodate connectors for cooling the CCDs using a thermo-electric cooler (TEC) (discussed in subsection \ref{sec:cooling}). The whole chamber is supported by four threaded rods onto an aluminum platform such that a liquid plate or a chiller system can be accommodated on the back of the bottom flange to remove the deposited heat by the TEC and control the hot plate temperature of the TEC.  

\begin{figure}
    \centering
    \begin{subfigure}{.33\textwidth}
    \includegraphics[width=\linewidth]{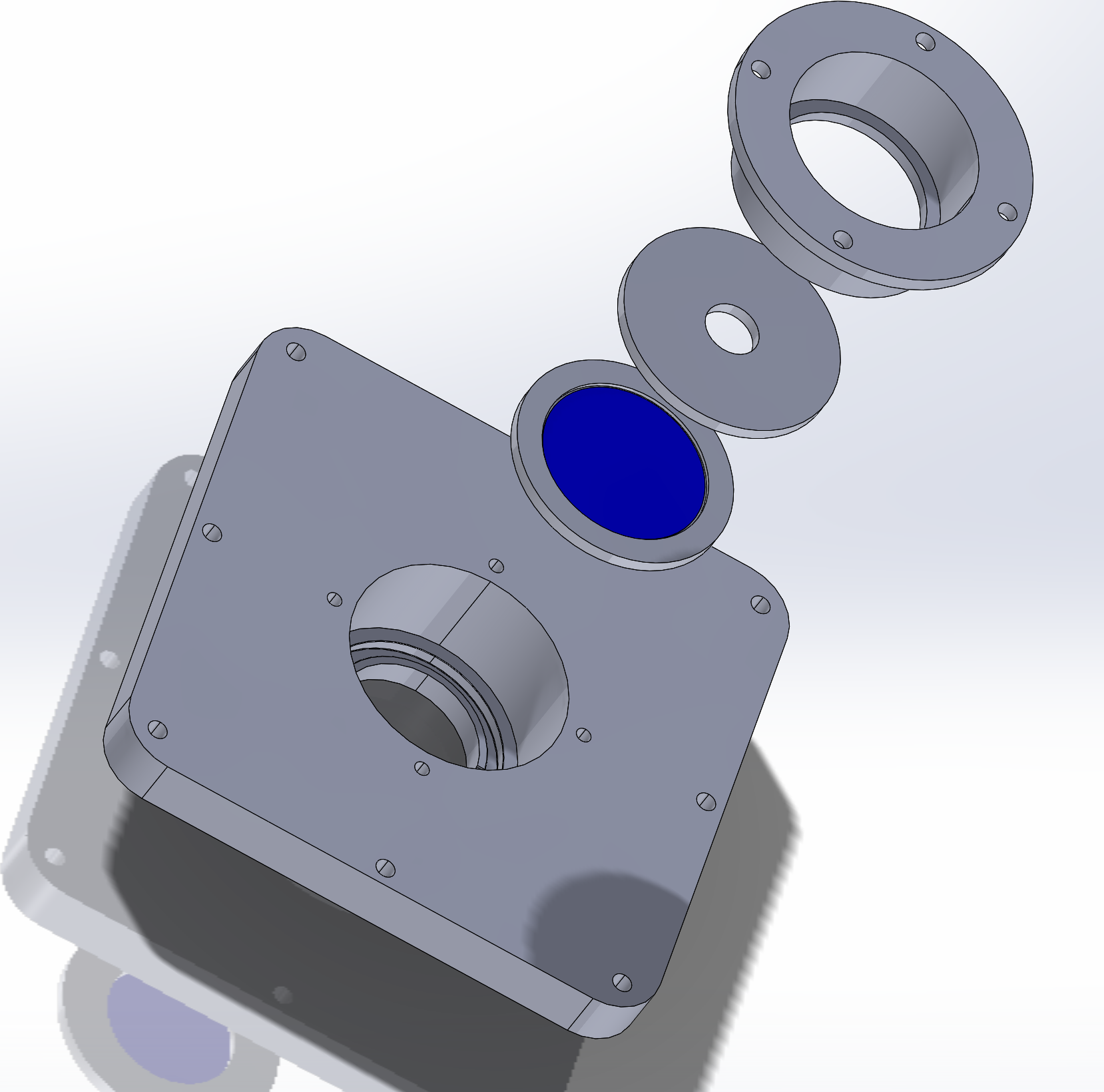}
    \caption{}
    \end{subfigure}
    \begin{subfigure}{.435\textwidth}
    \includegraphics[width=\linewidth]{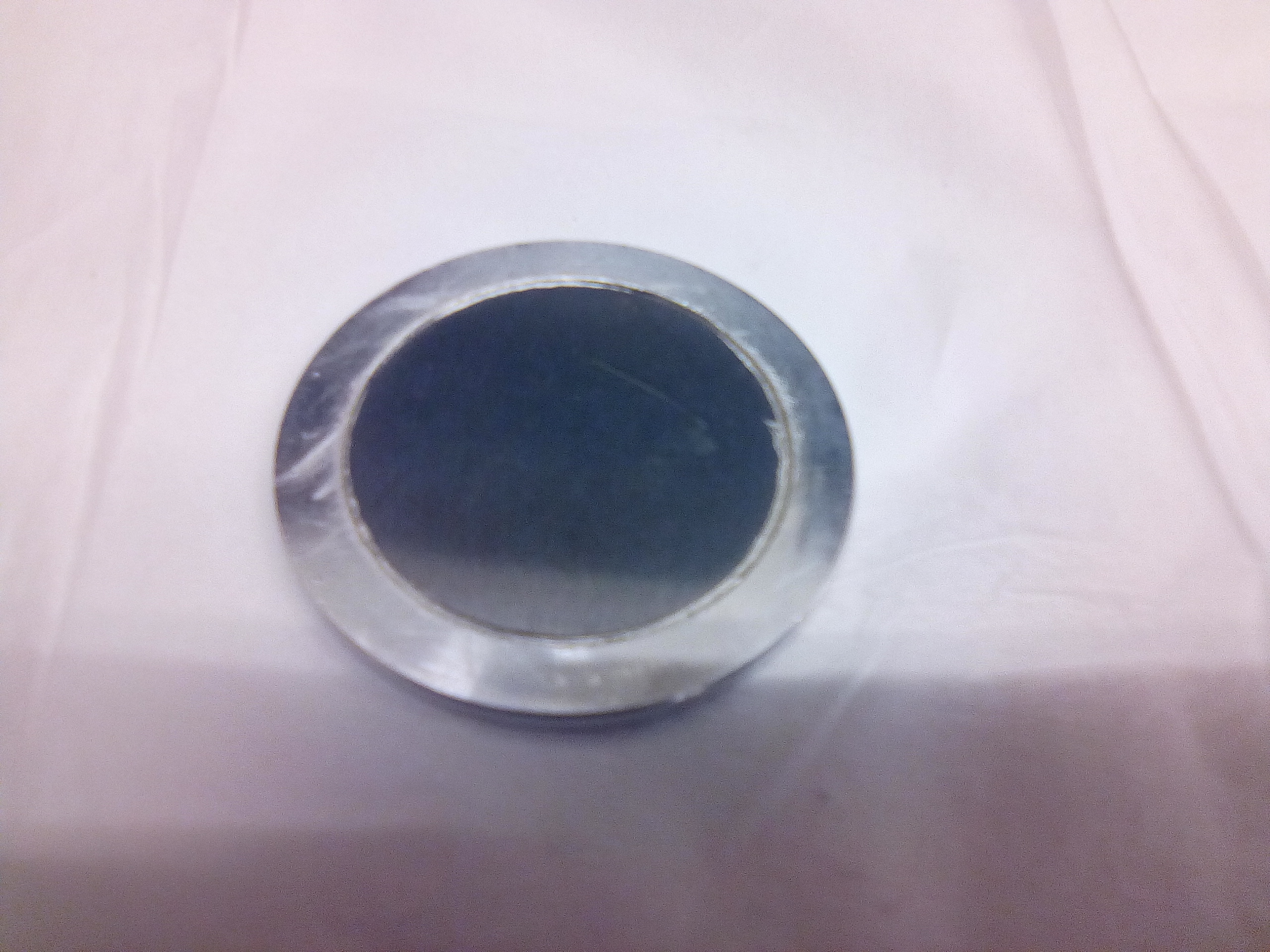}
    \caption{}
    \end{subfigure}
    \caption{(a) CAD model of the X-ray entrance window assembly along with the chamber top  flange. (b) The beryllium window: a 500 $\mu$m beryllium disc is epoxied on an aluminum mount.}
    \label{window}
\end{figure}

An X-ray entrance window is installed in the middle of the top flange of the chamber (see Fig. \ref{window}) directly above the detector.  
Fig. \ref{window}a shows the CAD model of the complete setup of the top flange, the beryllium X-ray window (shown in blue), and an aluminum seal.
A 500 $\mu$m thick beryllium disc of 3.8 cm diameter is epoxied on an aluminum mount (see Fig. \ref{window}b) which is clamped against the chamber top flange by an aluminum seal, as shown in Fig. \ref{window}(a). The window provides a $\sim$2.5 cm opening on the detector side to allow the X-rays to illuminate the detector active volume approximately uniformly. A 500 $\mu$m thick beryllium allows around 95 \% transmission of 5.9 keV photons, enabling effective use of the Mn k$_{\alpha}$ line at 5.9 keV from a standard $^{55}$Fe radioactive source. In the future we plan to install a thinner beryllium window in the chamber to allow for low energy characterization of the detectors.     

\subsubsection{Cooling and temperature control module:}
\label{sec:cooling}
\begin{figure}
    \centering
    \begin{subfigure}{.42\textwidth}
    \includegraphics[width=\linewidth]{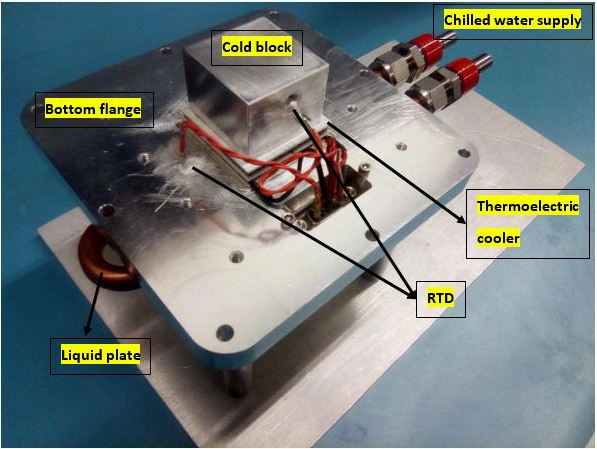}
    \caption{}
    \end{subfigure}\\
    \begin{subfigure}{.26\textwidth}
    \includegraphics[width=\linewidth]{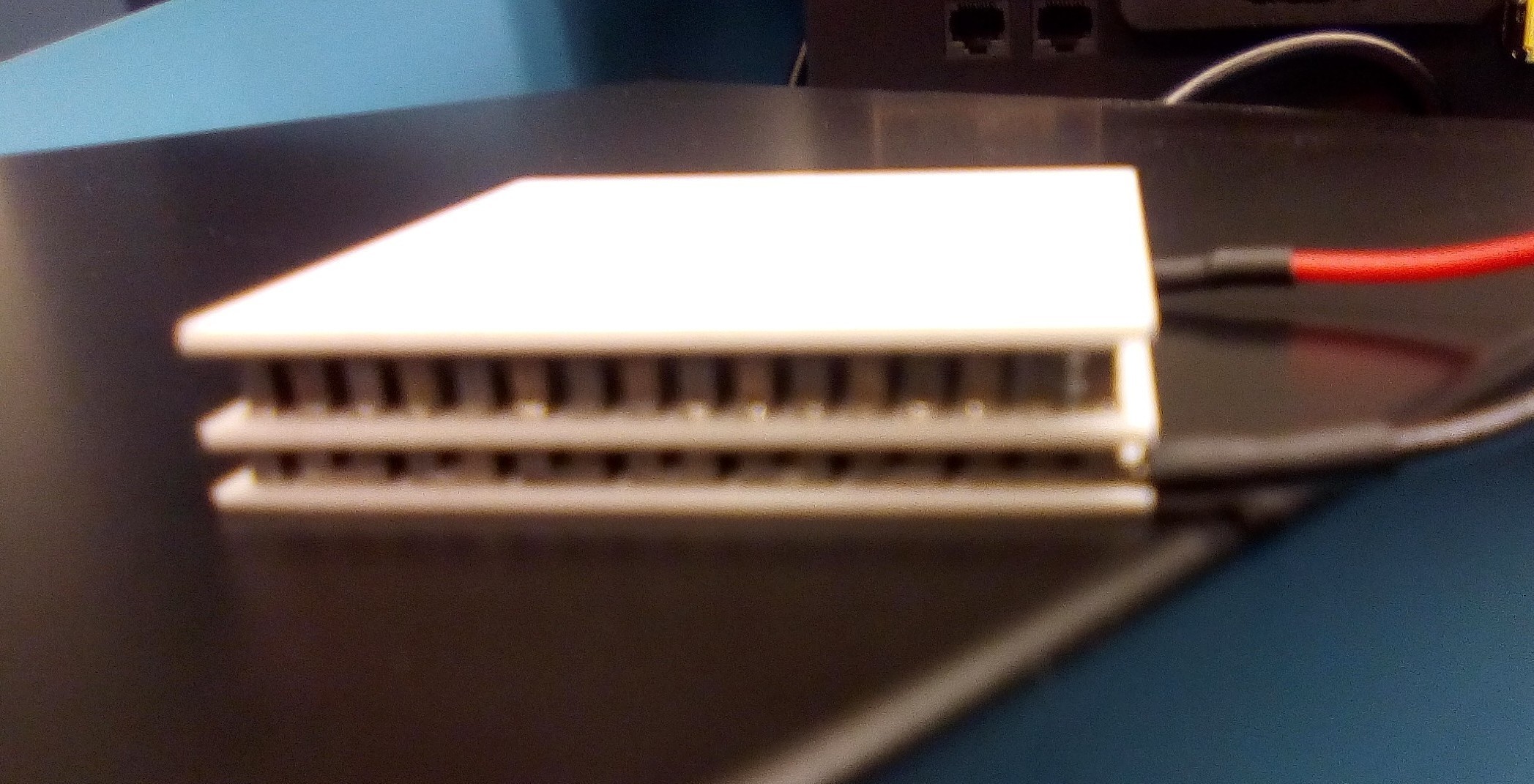}
    \caption{}
    \end{subfigure}
    \begin{subfigure}{.14\textwidth}
    \includegraphics[width=\linewidth]{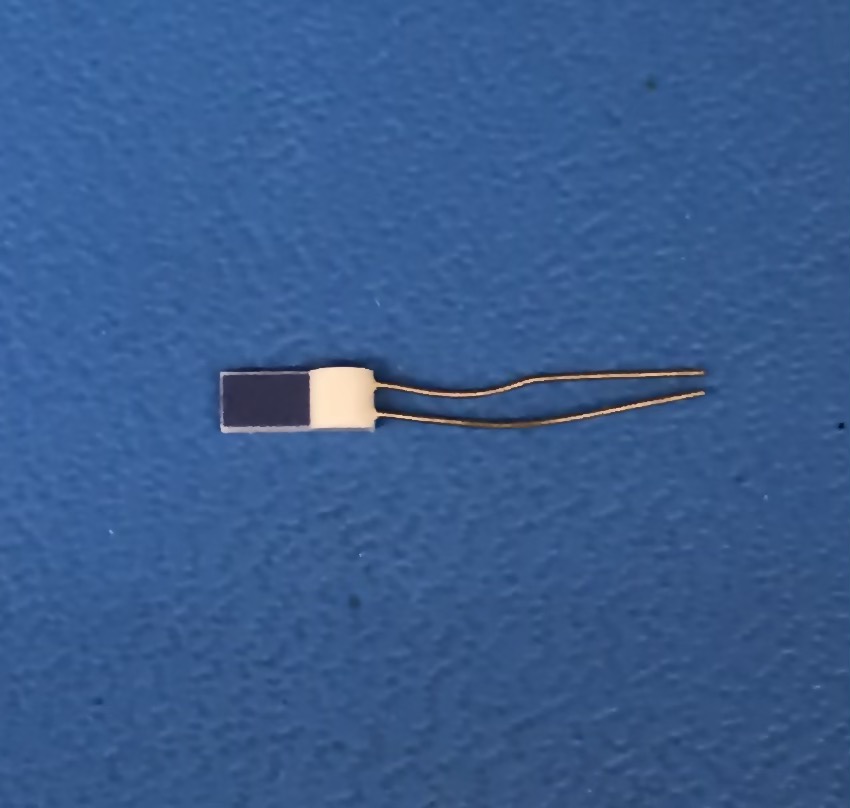}
    \caption{}
    \end{subfigure}
    \begin{subfigure}{.26\textwidth}
    \includegraphics[width=\linewidth]{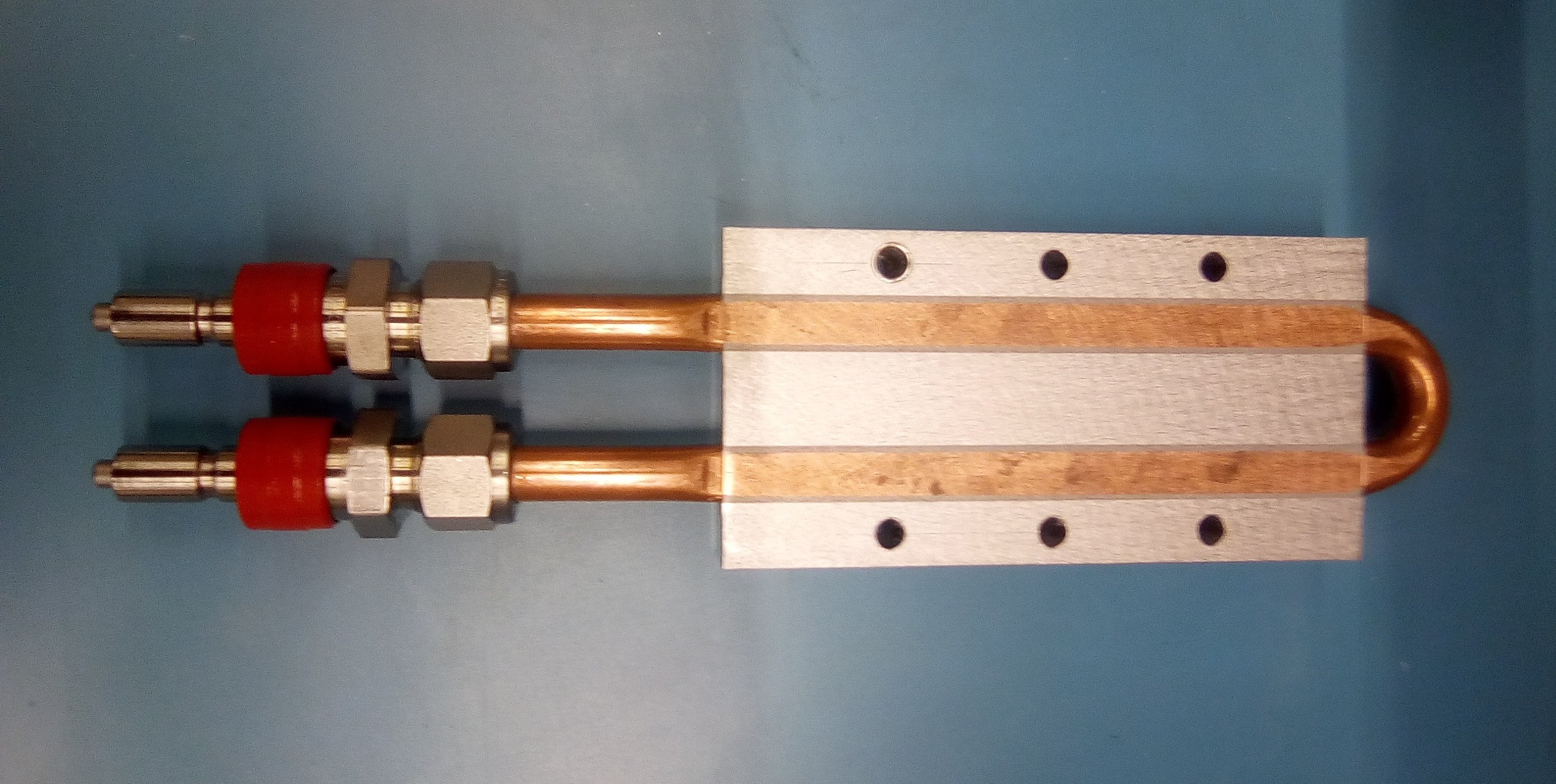}
    \caption{}
    \end{subfigure}
    \caption{(a) An inner view of the CCD chamber (Fig. \ref{chamber}) and the components required for cooling the detectors: $-$ a TEC (b) epoxied on the bottom flange, a cold block epoxied on the TEC, two RTDs (c) to measure temperatures from the cold and hot side of the TEC, and a liquid cooled metal plate (d) mounted on the back side of the bottom flange to dissipate heat generated by the TEC. A PID loop controls the current to the TEC, to stabilize the detector temperature.}
    \label{innerview}
\end{figure}

An important component of the characterization test stand is the cooling and control of the temperature of the CCDs. 
The cooling and temperature control module uses two resistance temperature detectors (RTDs), two Adafruit RTD readout boards, a thermoelectric cooler (TEC), a liquid-cooled aluminum block, and a Raspberry Pi controller to dissipate heat from the CCD and regulate temperature. 
The inner view of the chamber and the various components are shown in Fig. \ref{innerview}.
A 2-stage TEC (see Fig. \ref{innerview}b) from TE Technology\footnote{https://tetech.com/peltier-thermoelectric-cooler-modules/multi-stage/
} is mounted inside the chamber with hot side of the TEC directly epoxied (thermally conductive EPO$-$TEK H20E\footnote{http://bondingsource.com/h20e-epo-tek-epoxy}) on the bottom flange. 
The TEC operates at V$_{max}$ of 15.4 V and provides a substantial temperature difference between the hot and cold sides ($\Delta$T$_{max}$ = 84$^\circ$ for 30$^\circ$C hot plate temperature) with nominal operating conditions of the parameters. 
The aluminum cold block is mounted on the TEC cold side using the same epoxy.
The CCD lies on top of the aluminum cold block. We use thin copper foils between the back of the CCD package and the cold block for better thermal contact. 
With this arrangement, the CCD is in thermal contact with the cold side of the TEC through the cold block. By regulating the current through the TEC, the CCD is cooled down to low temperatures.
We use 2-wire Pt 1000 $\Omega$ RTDs (see Fig. \ref{innerview}c) from TEWA sensors LLC\footnote{https://www.digikey.com/en/products/detail/tewa-sensors-llc/TT-PT-1000B-2050-11-AUNI/9817197} for our temperature measurements.
One RTD is implanted near the top of the aluminum block using a thermally conductive and electrically insulating epoxy\footnote{https://epoxyinternational.com/thermalbond-195-thermally-conductive-electrically-insulating-compound-adhesive} to get the temperature reading from the cold side (we refer to this temperature as $T_{\mathrm{rtdCold}}$).
The hot side of the TEC is in thermal contact with the bottom metal flange, which dissipates heat to the liquid-cooled metal plate mounted on the back side of the flange. 
The liquid-cooled plate (see Fig. \ref{innerview}d), procured from D6 industry\footnote{https://d6industries.com/wp-content/uploads/2019/06/HBDS-02-4-PASS-0450.pdf}, is a compact,
high-performance precision-machined aluminum plate
with a continuous copper tube pressed into a
2 pass configuration. It provides a thermal resistance of 0.02$^\circ$C/Watt. The plate is connected to a chilled water supply flow to remove the dissipated heat.  
Another RTD is implanted on the bottom flange to enable temperature reading from the hot side of the TEC (refer to this temperature as $T_{\mathrm{rtdHot}}$). The temperature measurements from the RTDs are done by employing two Adafruit PT1000 RTD temperature sensors $-$ MAX31865
\footnote{https://www.adafruit.com/product/3648} (one for each RTD) and a Raspberry Pi controller to provide the necessary bias and clock signals to the temperature sensor.

An important requirement of the set up is to control the cold side temperature with high precision in order to avoid any temperature dependent fluctuations of CCD parameters (e.g gain, dark current etc.).
In this setup, the current to the TEC can be altered based on the temperature readings of the two RTDs to achieve the desired temperature.
The control is accomplished using a proportional-integral-derivative (PID) loop between the temperature signals and the power supply providing current to the TEC. However, the operation of the TEC based on the RTD temperature is not a linear relationship.
We instead use a series of modified equations~\citep{van2017practical} which use the difference between the true temperature and the set point along with a several setup dependent  measurable parameters in order to drive the required current to the TEC.
The modified equations boost the current above the predicted required amount for the TEC when the temperature is far from the set-point, while also limiting how large a current step can be taken between each sampling period. In this way, the desirable temperature range can be reached faster, but still at a safe rate. 

Equation \ref{eq:u}, which is the sum of equations \ref{eq:u1} to \ref{eq:u3}, determines the PID term which is fed into equation \ref{eq:current}. 
\begin{equation}
    \mathrm{error} = T_{\mathrm{rtdCold}} - T_{\mathrm{set}}
\end{equation}
\begin{equation}
\label{eq:u1}
    u_1 = \frac{-1}{t_c}\times \mathrm{error}_{\mathrm{current}}
\end{equation}
\begin{equation}
\label{eq:u2}
    u_2 = -K_I \times \sum_{n=t-20}^t \mathrm{error}_n
\end{equation}
\begin{equation}
\label{eq:u3}
    u_3 = -K_m \times (T_{\mathrm{rtdHot}} - T_{\mathrm{rtdCold}})
\end{equation}

\begin{equation}
\label{eq:u}
    u = u_1 + u_2 + u_3
\end{equation}
Equation \ref{eq:current} then calculates the final driving current needed to be delivered to the TEC, assuming such a current does not exceed the previously mentioned limits.

\begin{equation}
    \label{eq:current}
    \mathrm{current} =  \begin{cases}
    \alpha \times \left(\frac{S_m \times T_{\mathrm{rtdCold}}}{R_m}-\sqrt{\frac{S^2_m\times T^2_{\mathrm{rtdCold}}}{R^2_m}-\frac{u}{0.5\times R_m}}\right) + \frac{\mathrm{error}^{p_1}}{\beta}+\frac{\mathrm{error}^{p_2}}{\gamma} & u \geq -\frac{S^2_m\times T^2_{\mathrm{rtdCold}}}{R^2_m} \\
     \alpha \times \left(\frac{S_m \times T_{\mathrm{rtdCold}}}{R_m}\right) + \frac{\mathrm{error}^{p_1}}{\beta}+\frac{\mathrm{error}^{p_2}}{\gamma} & u < -\frac{S^2_m\times T^2_{\mathrm{rtdCold}}}{R^2_m}
    \end{cases} 
\end{equation}
Here $T_{\mathrm{rtdCold}}$ is the temperature of the RTD nearest to the CCD, error$_\mathrm{current}$ is the error on the current time acquisition step, and $T_{\mathrm{rtdHot}}$ is the temperature of the RTD on the metal flange. The remaining quantities are parameters tuned during testing and are specific to the system being used. The higher order error terms in the equation for current are boost factors for when the error is high, but do not play a significant role when the measured temperature gets close to the set-point.

\begin{figure}
    \centering
    \includegraphics[width=0.7\linewidth]{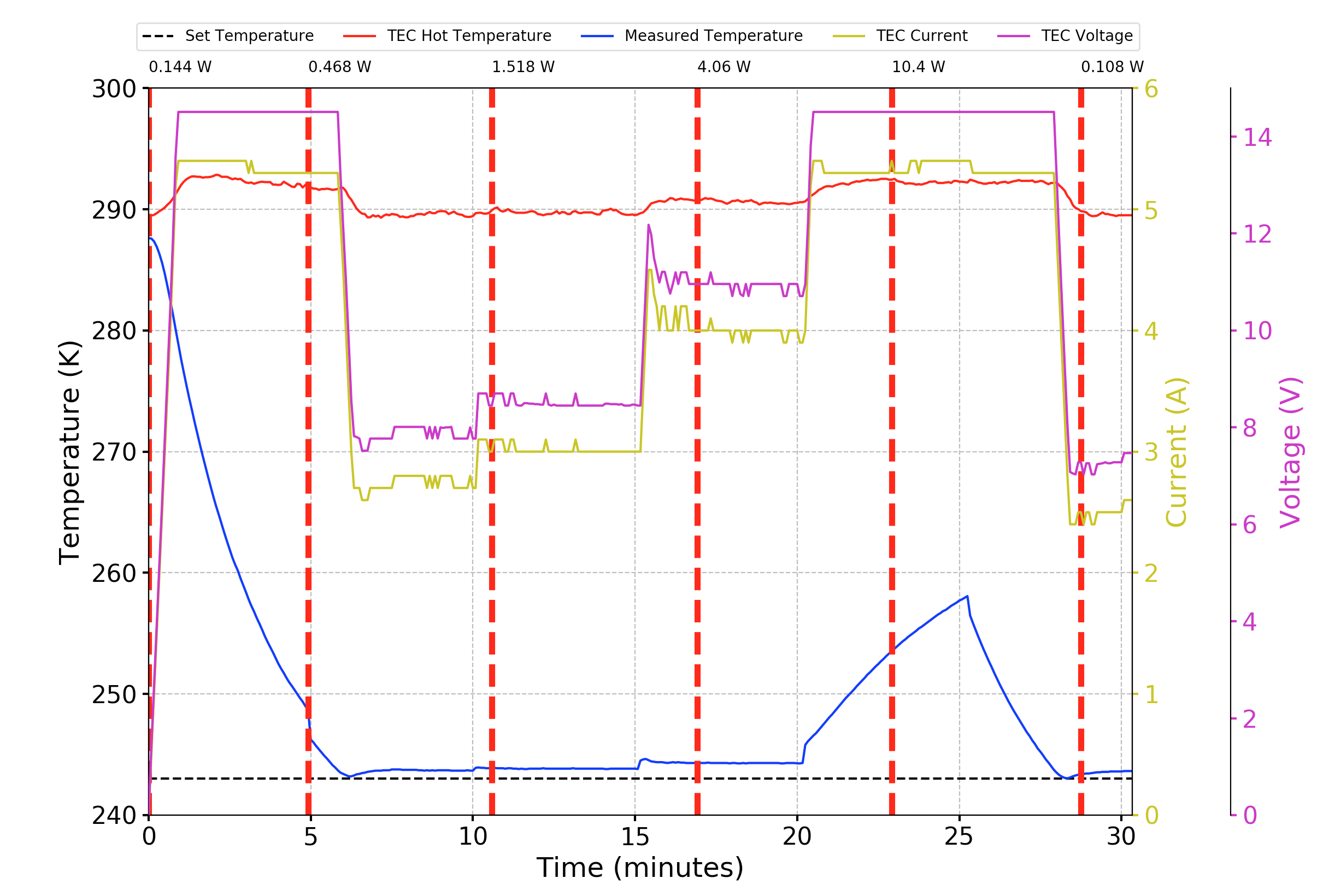}
    \caption{Testing of the cooling module for various known amounts of heat deposition on the cold block. For a set temperature of 243 K (black dashed line), the blue line tracks the cold side temperature. We see the cooling module is capable of cooling down to 243 K for heat depositions $<$5 Watts. The solid red line indicates the temperature of the hot side of the TEC. The yellow and violet lines show the changes in the TEC current and voltage with heat deposition. The vertical dashed red lines indicate the time at which the deposited heat is changed, with the power shown as a label above the lines.}
    \label{cooling_expt}
\end{figure}

As a proof of concept, we mounted a small heating element on the cold block to deposit a known amount of heat on the cold block. Fig. \ref{cooling_expt} shows the performance of the cooling module for various known heat depositions and required TEC parameter values to achieve the desired cooling.  
We set the cold side temperature to 243 K, shown by the dashed black line. The solid blue and the red lines track the changes in the cold side and hot side temperatures, respectively, with changes in the heat deposition. The changes in heat deposition are demarcated by the vertical dashed red lines with the power shown as a label at the top. We see that the cooling module is capable of cooling down the CCDs to temperature 243 K for heat depositions of $<$5 Watts for nominal operating conditions of the TEC parameters like  current and voltage (shown in yellow and violet respectively). We also estimate the stability in the cold side temperature to be around $\pm$0.2$^\circ$ around the set point, which demonstrates the high efficacy of the PID loop. In actual CCD operations (particularly for the CCDs from MITL CCID85 series), we expect smaller heat depositions, which should enable cooling of the detectors down to even lower temperatures. 


\subsection{Overview of the X-ray CCDs (CCID85 devices) and readout module}
\label{readout}

In this section, we will first give an overview of the X-ray devices that we are testing, followed by a description of the readout module, which consists of a preamplifier board and an Archon controller. The preamplifier board reads and amplifies the analog output of the CCDs ,whereas the Archon controller provides the necessary bias and clock signals to run them, digitize the output signal and perform correlated double sampling (CDS) to generate 2D images. 

\subsubsection{CCID85 X-ray CCDs}
For our experiment, we use the prototype X-ray CCDs (CCID85 detectors) being developed by MIT Lincoln laboratory. These detectors are fabricated in an n-channel, low-voltage, single-poly process and use fast, low-noise amplifiers. These devices have the potential to meet the frame rate and noise performance requirements of the next generation X-ray astronomy missions.

\begin{figure}
    \centering
    \begin{subfigure}{.36\textwidth}
    \includegraphics[width=\linewidth]{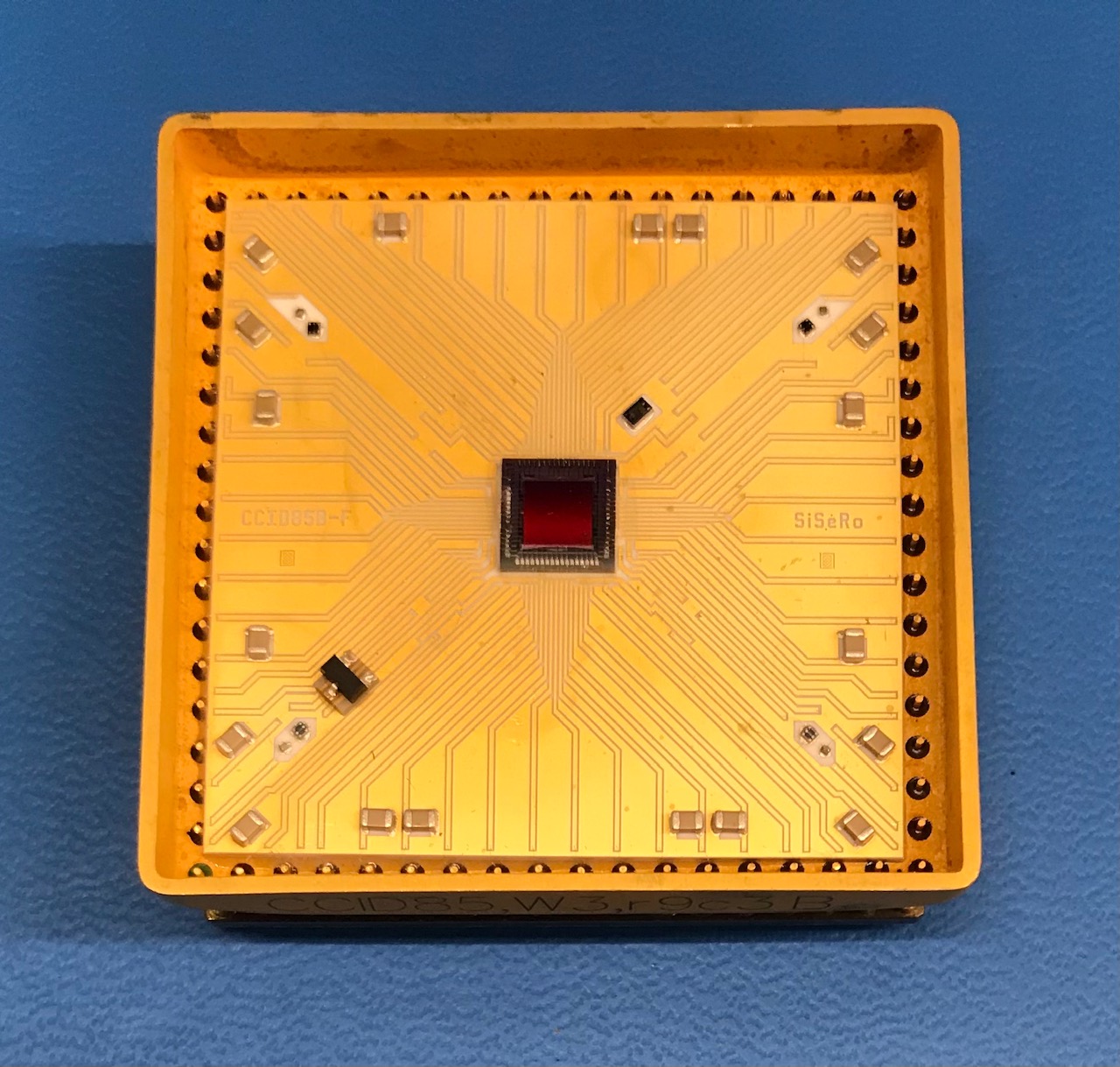}
    \caption{}
    \end{subfigure}
    \begin{subfigure}{.33\textwidth}
  \includegraphics[width=\linewidth]{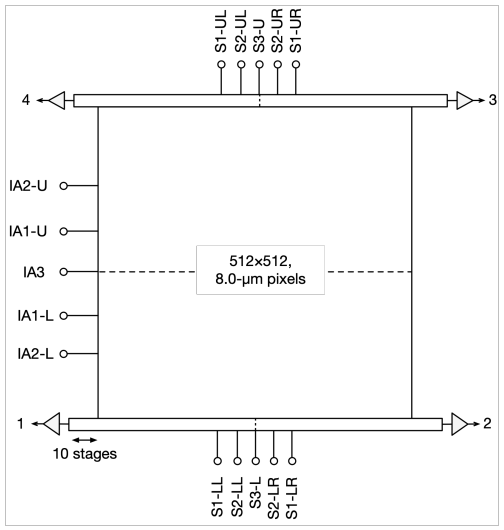}
  \caption{}
  \end{subfigure}
  \begin{subfigure}{.39\textwidth}
     \includegraphics[width=\linewidth]{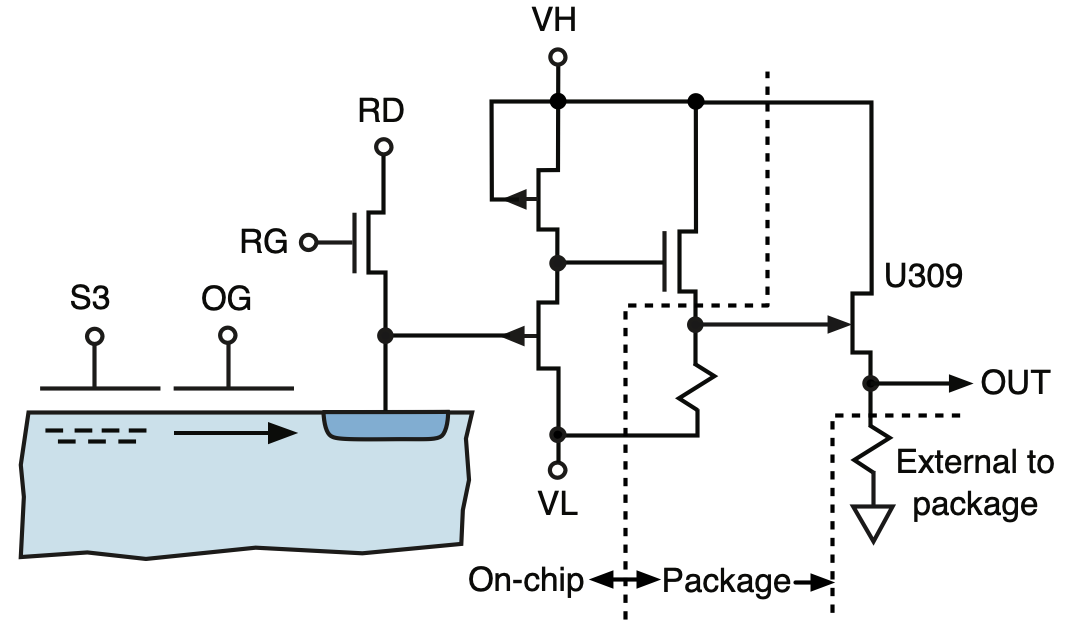}
    \caption{} \label{fig:1a}
  \end{subfigure}%
   \begin{subfigure}{.38\textwidth}
    \includegraphics[width=\linewidth]{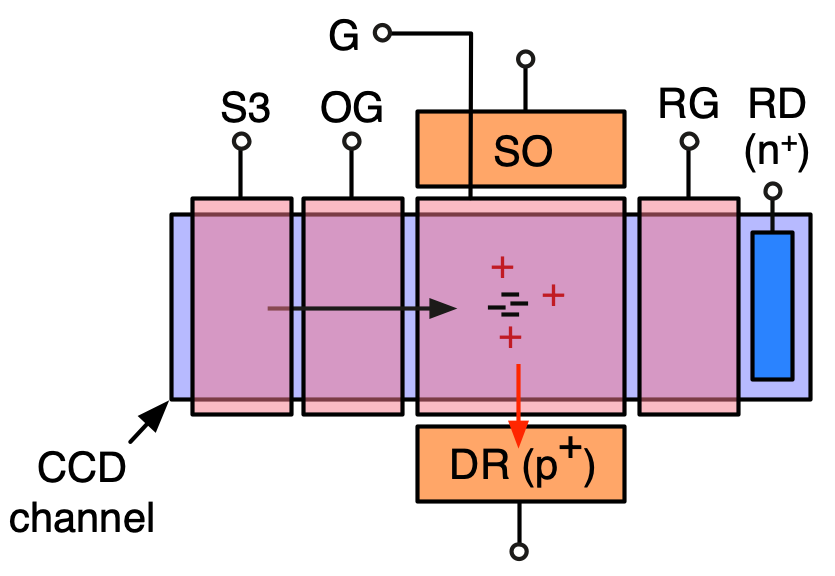}
    \caption{} \label{fig:1a}
  \end{subfigure}
    \caption{CCID85 X-ray CCDs from MIT Lincoln laboratory. (a) One CCID85 sensor in its package and (b) its schematic. The devices are equipped with 2 types of amplifiers $-$ (c) P-JFET floating diffusion amplifier and (d) Single Electron Sensitive Readout (SiSeRO) floating gate amplifier. See text for more details.}
    \label{ccid85}
\end{figure}

Fig. \ref{ccid85}a shows one CCID85 device package. The sensitive volume is $\sim$4 mm $\times$ 4 mm in size with 512 $\times$ 512 array of 8 $\mu$m pixels. 
Currently, the CCID85 detectors are front-illuminated devices. Fig. \ref{ccid85}b displays the schematic of the imaging and serial clocking gates. Each device is equipped with 4 amplifier nodes at the 4 corners of the device.    
For testing flexibility, both the serial and imaging gates are designed so that the entire device can be read out of any of the four amplifiers or operated as a four-port device.

There are two classes of amplifiers in the CCID85 devices $-$ P-channel JFET amplifiers and Single Electron Sensitive Readout (SiSeRO) amplifiers. 
Unlike conventional CCDs which use buried-channel MOSFETs for the sense transistor, the first one in these devices uses a P-JFET source follower for the first stage followed by a buried-channel MOSFET as a second stage to increase the bandwidth of the amplifier. There can also be an optional third stage of U309 N-JFET source follower in the package (see Fig. \ref{ccid85}c). 
The SiSeRO amplifier, shown in Fig. \ref{ccid85}d, on the other hand, is a floating-gate amplifier, where a P-MOSFET gate sits on top of the CCD channel. When a CCD charge packet is transferred beneath the gate, it modulates the drain current of the transistor, which is proportional to the source signal strength. The signal is brought directly out of the package through the source or drain of the transistor, enabling the device to be operated in a drain-current readout mode. The advantage of drain (or current) readout is that, unlike P-JFET amplifiers, the readout speed is not limited by the RC time time constant of the following readout circuit, where `R' is the reciprocal of the transistor transconductance (g$_m$) and `C' is input capacitance of the amplifiers plus the various parasitic capacitances in the circuit. SiSeRO amplifiers are therefore anticipated to provide multifold improvements in readout speed and  noise performance (see \citep{bautz19} for more details). To establish the functionality of our test stand, we first characterized the performance of the P-JFET amplifiers. Characterization of the SiSeRO amplifiers will be discussed elsewhere. 
In the next subsection, we describe the preamplifier board as the first component of the readout module.  

\subsubsection{Preamplifier board}
\label{preamp}

\begin{figure}
    \centering
    \includegraphics[width=0.5\linewidth]{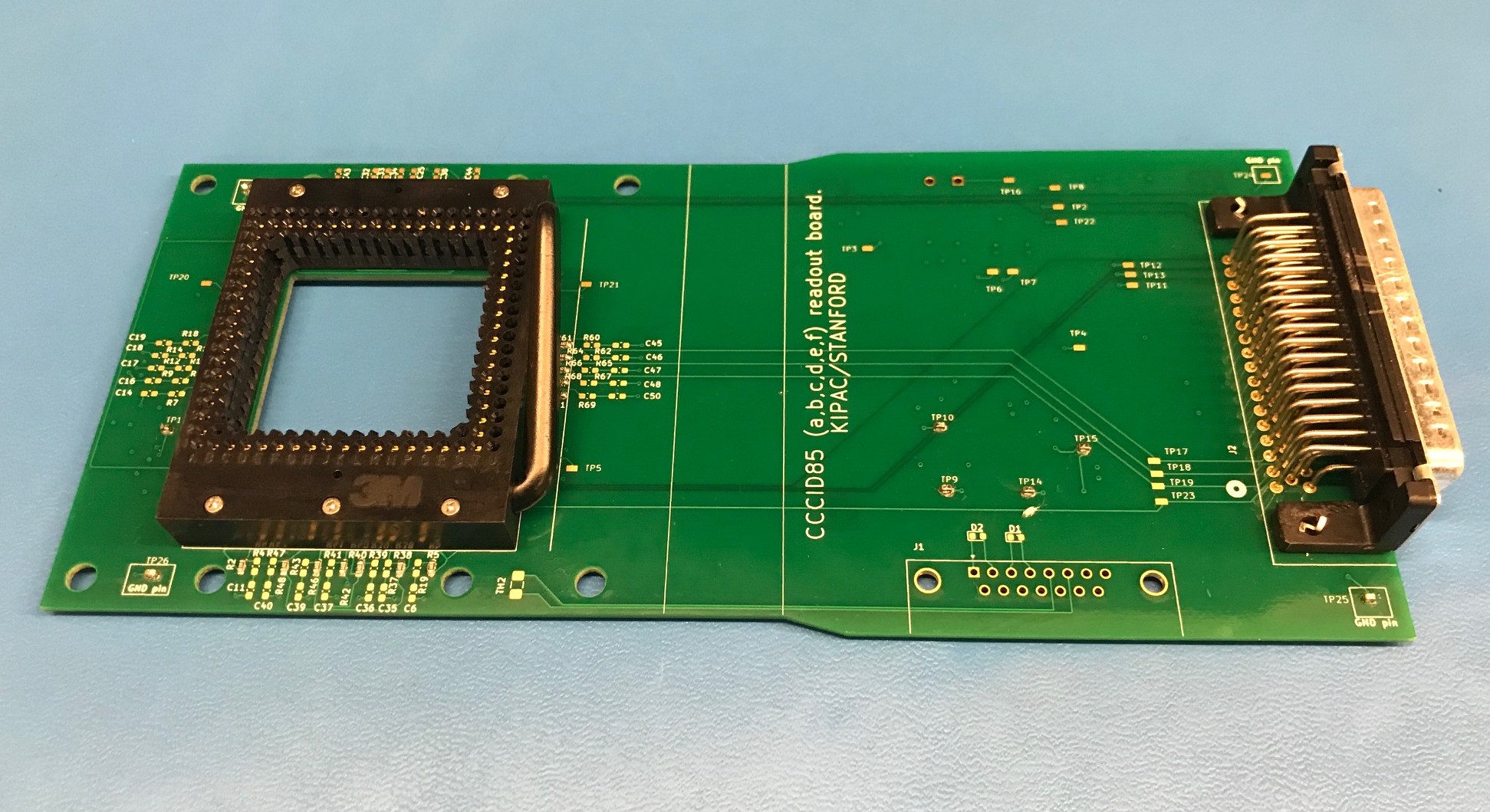}
    \caption{Preamplifier board designed at Stanford to readout the CCID85 X-ray CCDs devices.}
    \label{preamp_board}
\end{figure}

Fig. \ref{preamp_board} shows the preamplifier board that we have designed for our experiment. 
It is a compact (20 cm $\times$ 9 cm) board with 6 conductive layers. In the final assembly of ``tiny-box," the preamplifier board is epoxied (potted) in the side flange of the CCD chamber (see Fig. \ref{chamber}). The section of the board confined by the two vertical lines, as shown in the figure, is the potting region. The left part of the board with the CCD mounting socket is placed inside the chamber.
The cold block, shown in Fig. \ref{innerview}a), fits through the square hole in the board.
We use a 361 (19 $\times$ 19) position PGA ZIF socket procured from 3M, machined to mount the MITL CCID85 devices (68 output pins in total), such that the backside of the device package is in thermal contact with the cold block. As mentioned above, we use thin copper foils between the two surfaces for better heat transfer. A 50-pin connector, on the other side of the board, interfaces the preamplifier board with the Archon controller.    

Here we briefly summarize the various functions that the preamplifier board performs.
\begin{enumerate}
    \item It works as an interface to provide bias signals to run the CCDs and clock signals for transfer of charge in the devices.
    \item It provides an RC filtering circuit for the clocks and noise filtering of the DC bias signals.
    \item The board reads out the CCD output signals (both from P-JFET and SiSeRo amplifier output), amplifies and converts to fully differential output to interface with the differential ADCs.
    \item The board also accommodates two RTDs to provide temperature measurements of both the board and CCD from AD590 sensors\footnote{https://www.analog.com/en/products/ad590.html$\#$product-overview} installed inside the CCD package.   
\end{enumerate}
The 3$^{rd}$ entry in the list is critical to ensure that the readout speed of the CCD on-chip amplifiers is preserved during the amplification stage, and the contribution of the readout module to the total noise is as low as possible. 

As we discuss above, there are two distinct on-chip amplifiers in the CCD devices $-$ P-JFET and SiSeRO. The preamplifier circuit is designed to accommodate both types of amplifiers in the same preamplifier board.   
For the P-JFET amplifier output, we considered a number of different configurations, performing simulations in LTspice for each configuration. Because we are interested in high readout speed, and the fact that noise increases with the overall bandwidth of the circuit ($\sqrt{BW}$), this imposes restrictions on the choice of individual amplifiers. 
For the CCD on-chip amplifier, a transconductance (g$_m$) of 166 $\mu$S (or R=1/g$_m$ = 6 k$\Omega$) results in a CCD intrinsic noise around 10 nV/$\sqrt{Hz}$ (noise spectral density, S$_v$(f) $=$ $\sqrt{4KTR}$ nV/$\sqrt{Hz}$) at room temperature. A similar voltage noise at the input of the readout module is desired to have an overall low noise from the system.      
For simulations, we therefore considered amplifiers with 1. low voltage and current noise, 2. large bandwidth, 3. lower input capacitance such that the RC time constant of the circuit is low (`C' is the total input capacitance of the circuit and `R' is the reciprocal of transconductance (g$_m$) of the CCD on-chip amplifier). 

\begin{figure}
    \centering
    \includegraphics[width=1.\linewidth]{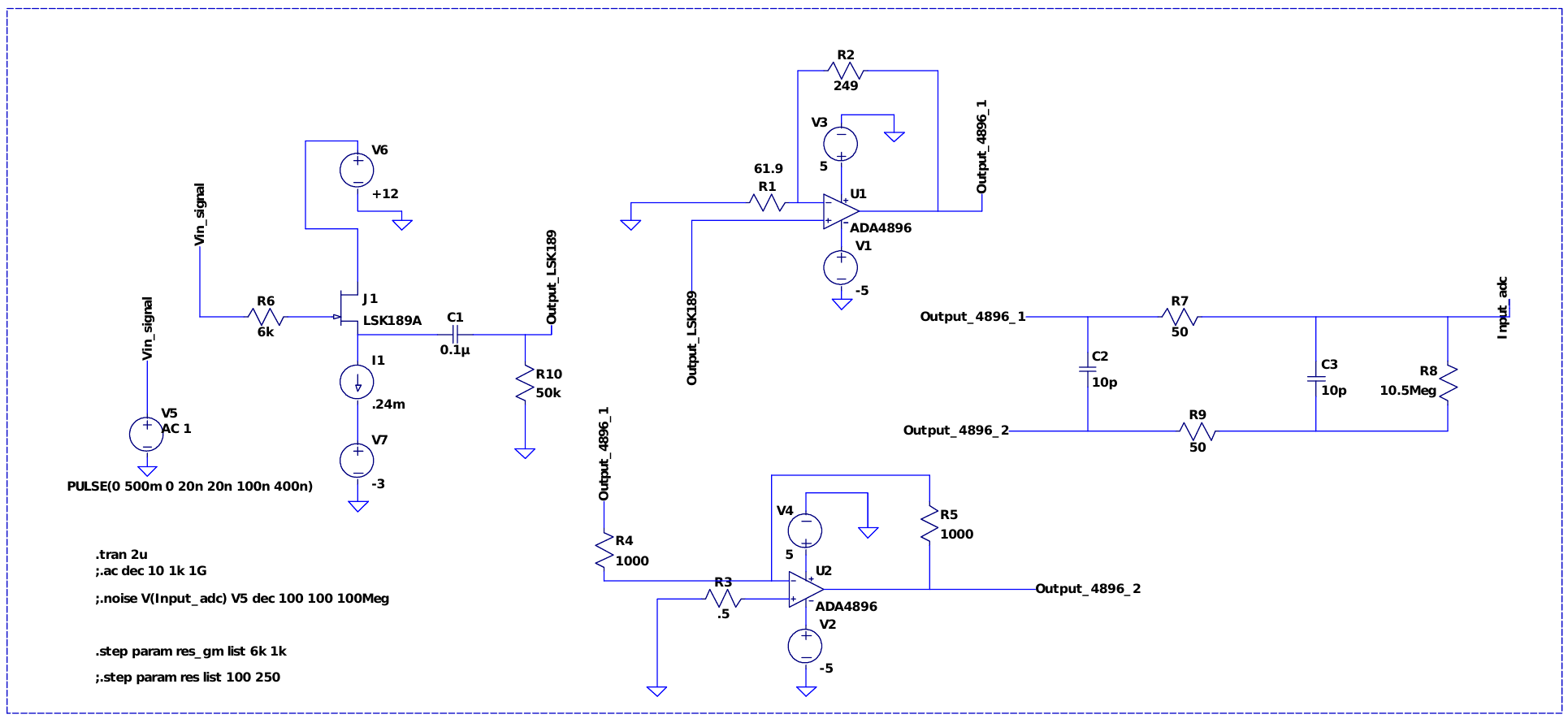}
    \includegraphics[width=1.\linewidth]{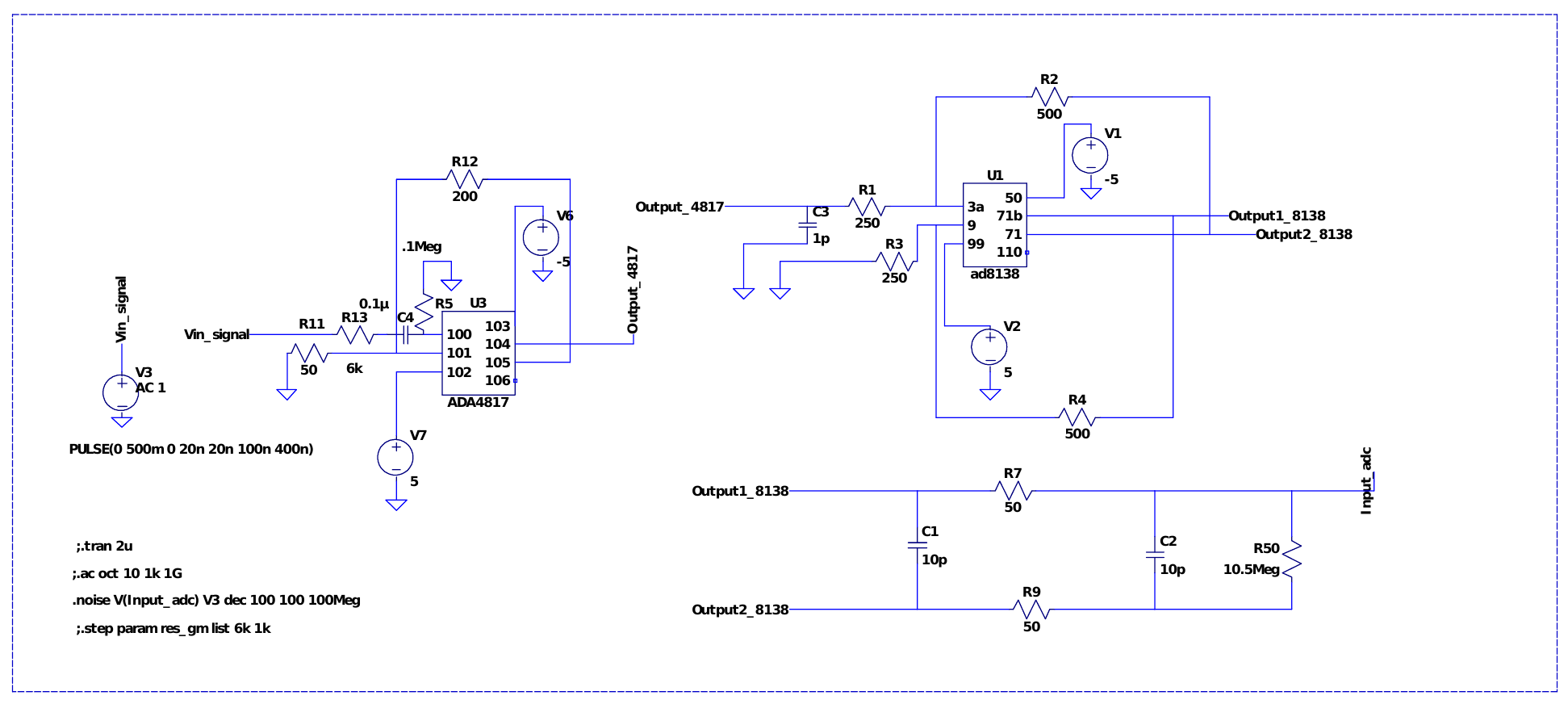}
    \caption{Schematic of the two different preamplifier chains implemented in the preamplifier board (Fig. \ref{preamp_board}) to read out and amplify the CCD P-JFET output. Top: preamplifier configuration (C1) $-$ an N-JFET source follower at the first stage is followed by two opamps to generate a fully differential signal at the output. Bottom: Preamplifier configuration (C2) $-$ a single ended opamp at the first stage is followed by a fully differential ADC driver.}
    \label{preamp_schematic}
\end{figure}

Out of all the configurations simulated, two different configurations show promising results in terms of large bandwidth and low noise, schematics of which are shown in Fig. \ref{preamp_schematic}.
The first configuration uses a source follower (N-channel LSK189 amplifier from Linear Systems\footnote{http://www.linearsystems.com/product-search-result.html?type=products\&partnumber=LSK189}) at the first stage followed by two single-ended operational amplifiers (ADA4896 opamps from Analog Devices\footnote{https://www.analog.com/en/products/ada4896-2.html$\#$product-overview}). The source follower drives the CCD output to the non-inverting input of one ADA4896 opamp. Output of this stage is fed to the inverting input of the other ADA4896 amplifier. Gains of the two opamps are optimized to produce a fully differential signal at the output. Both LSK189 and ADA4896 are low noise ($<$2 nV / $\sqrt{Hz}$), high bandwidth, low power amplifiers. The 6 k$\Omega$ resistor (R6 in the schematic) represents the transconductance of the on-chip P-JFET+MOSFET amplifier.   
The second configuration uses a single ended opamp (ADA4817 from Analog Devices\footnote{https://www.analog.com/en/products/ada4817-1.html$\#$product-overview}) at the first stage instead of a source follower. In this configuration, the overall bandwidth critically depends on the RC time constant as the 6 k$\Omega$ resistor is directly coupled to the non-inverting input of the amplifier. ADA4817 provides extremely low input capacitance of $\sim$1.3 pF compared to other amplifiers and therefore is best suited for this type of configuration. 
The low input capacitance and high speed of the amplifier results in a large bandwidth for this configuration, particularly when the input resistor (1/g$_m$) is small.  
At the second stage, a differential ADC driver, AD8138 from Analog Devices,\footnote{https://www.analog.com/en/products/ad8138.html$\#$product-overview} produces a fully differential signal at the input of ADCs. The overall speed of the circuit is mostly independent of the second stage. ADA4817 and AD8138 have slightly higher noise ($<$5 nV / $\sqrt{Hz}$) compared to the components considered in the first configuration. Therefore we expect better noise performance from the first configuration (hereafter C1) whereas the second configuration (hereafter C2) is more suitable for higher readout speeds.  

\begin{figure}
    \centering
    \begin{subfigure}{.42\textwidth}
    \includegraphics[width=\linewidth]{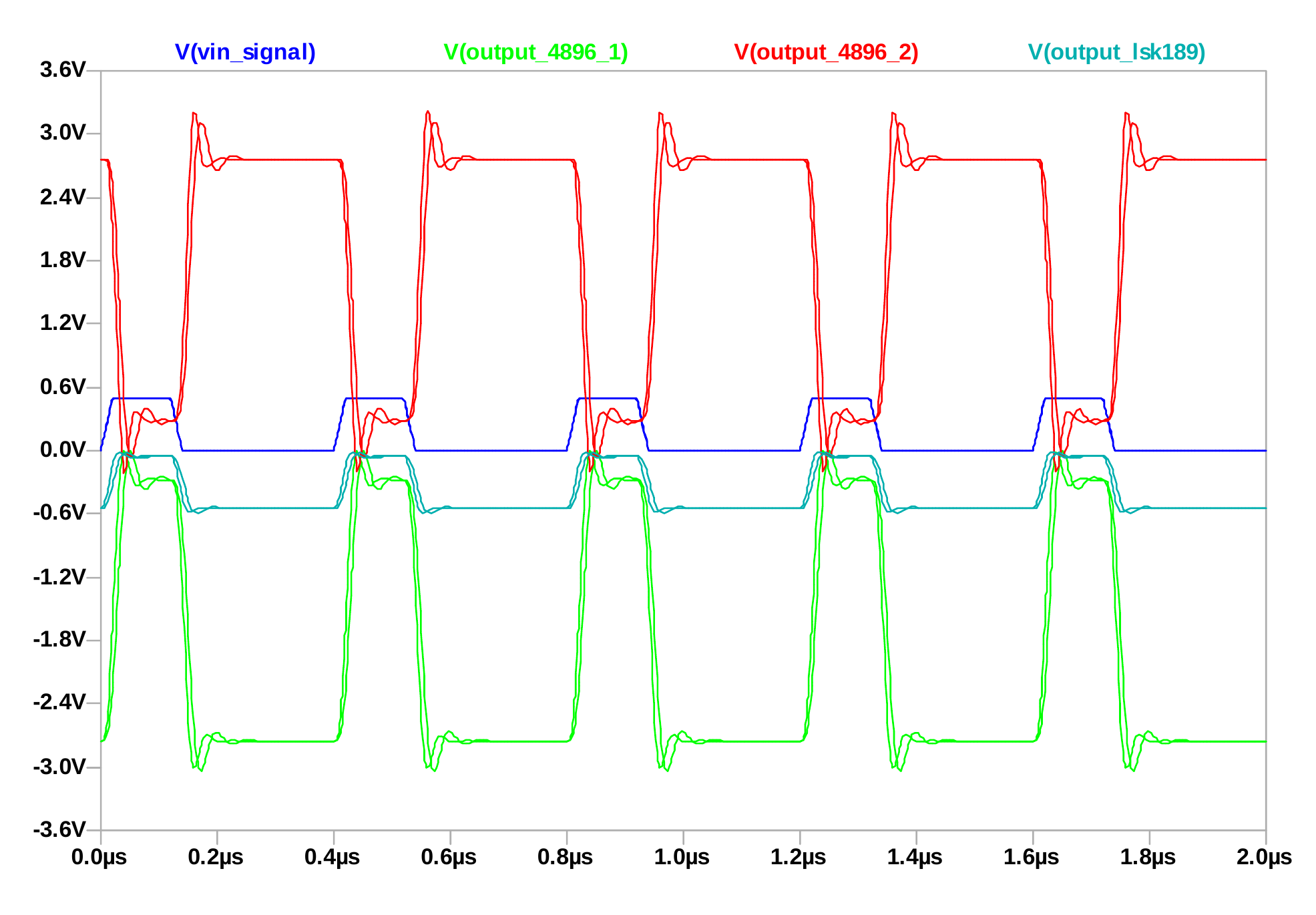}
    \caption{}
    \end{subfigure}
    \begin{subfigure}{.42\textwidth}
    \includegraphics[width=\linewidth]{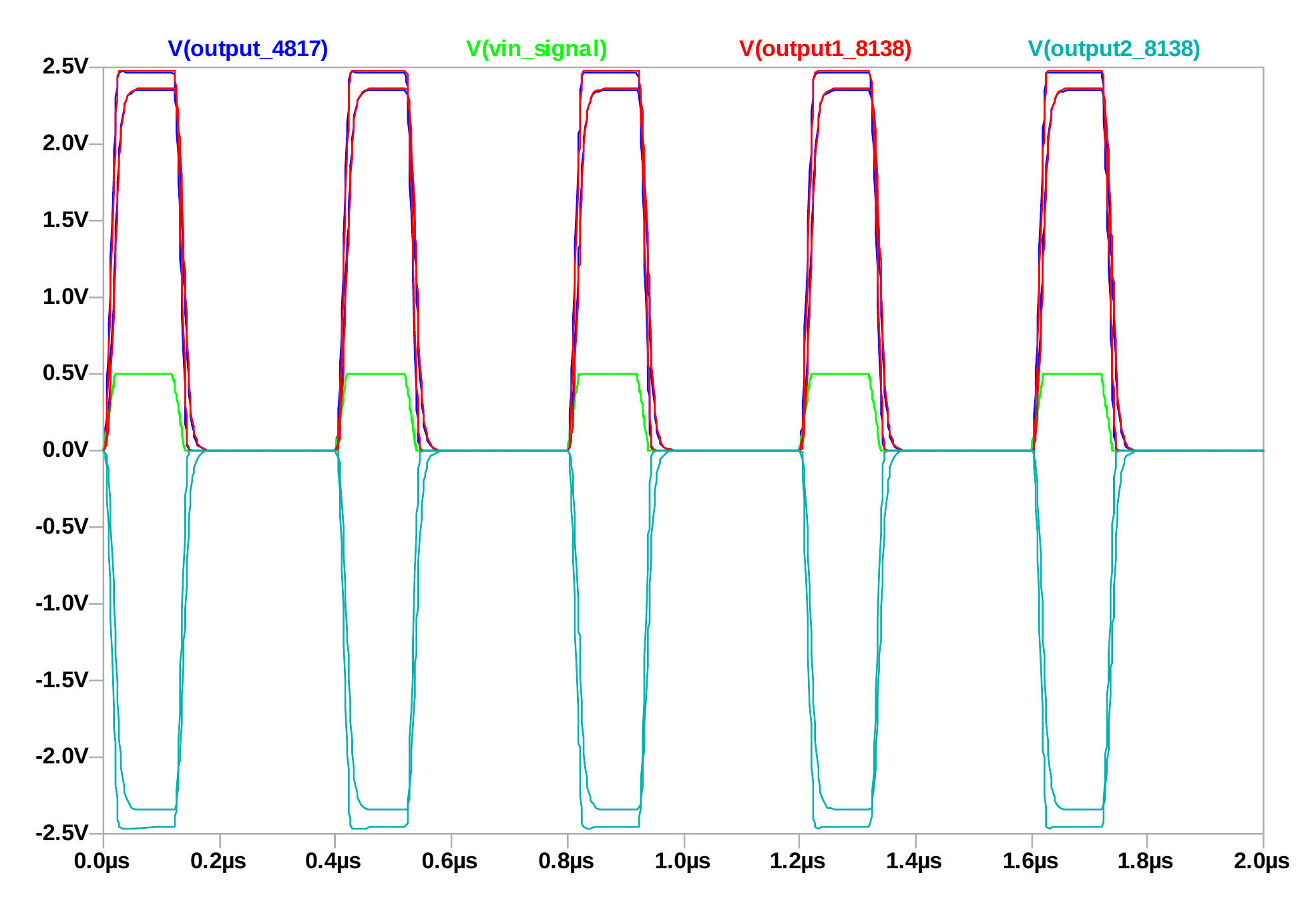}
    \caption{}
    \end{subfigure}
    \begin{subfigure}{.42\textwidth}
    \includegraphics[width=1.05\linewidth]{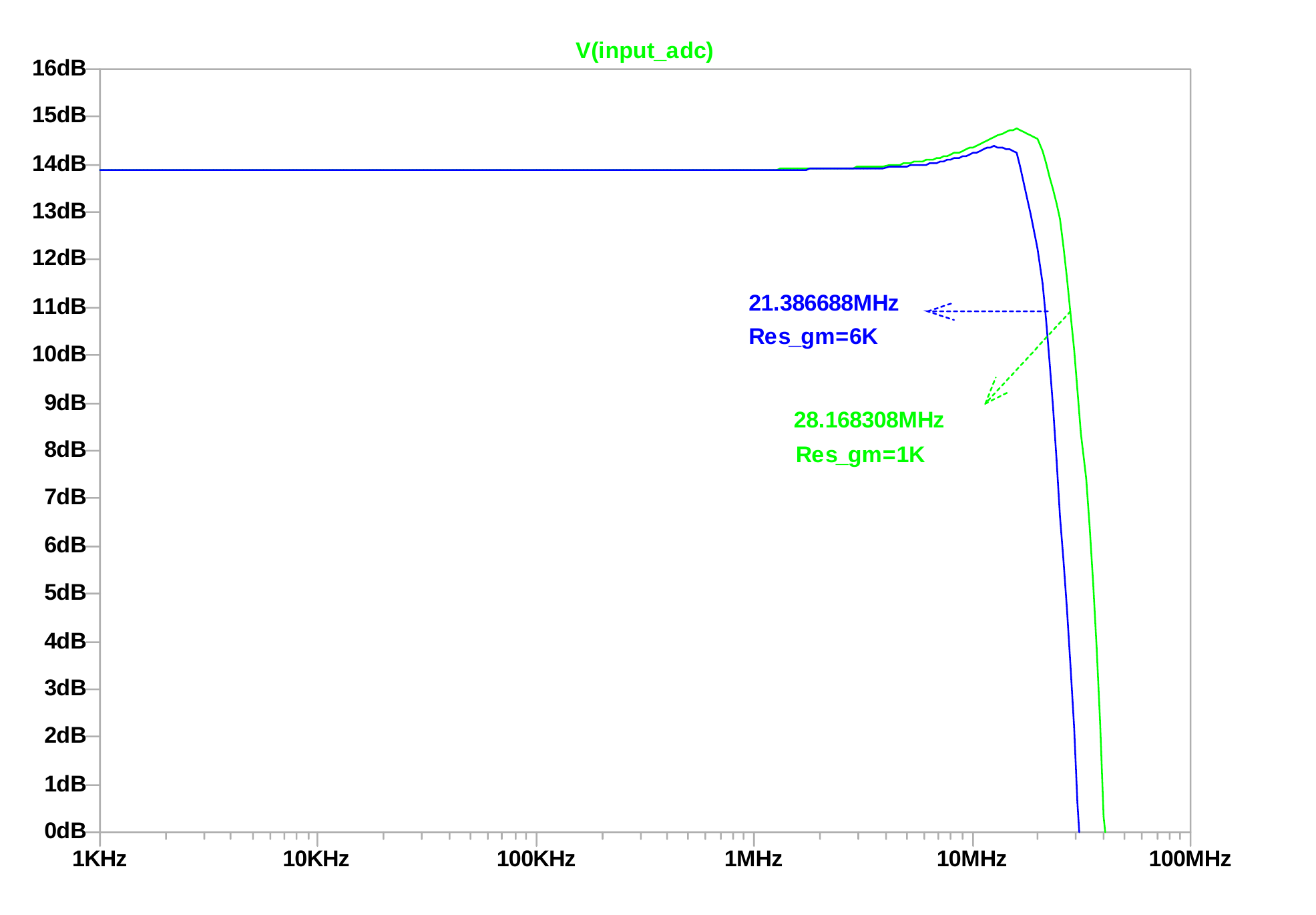}
    \caption{}
    \end{subfigure}
    \begin{subfigure}{.42\textwidth}
    \includegraphics[width=1.05\linewidth]{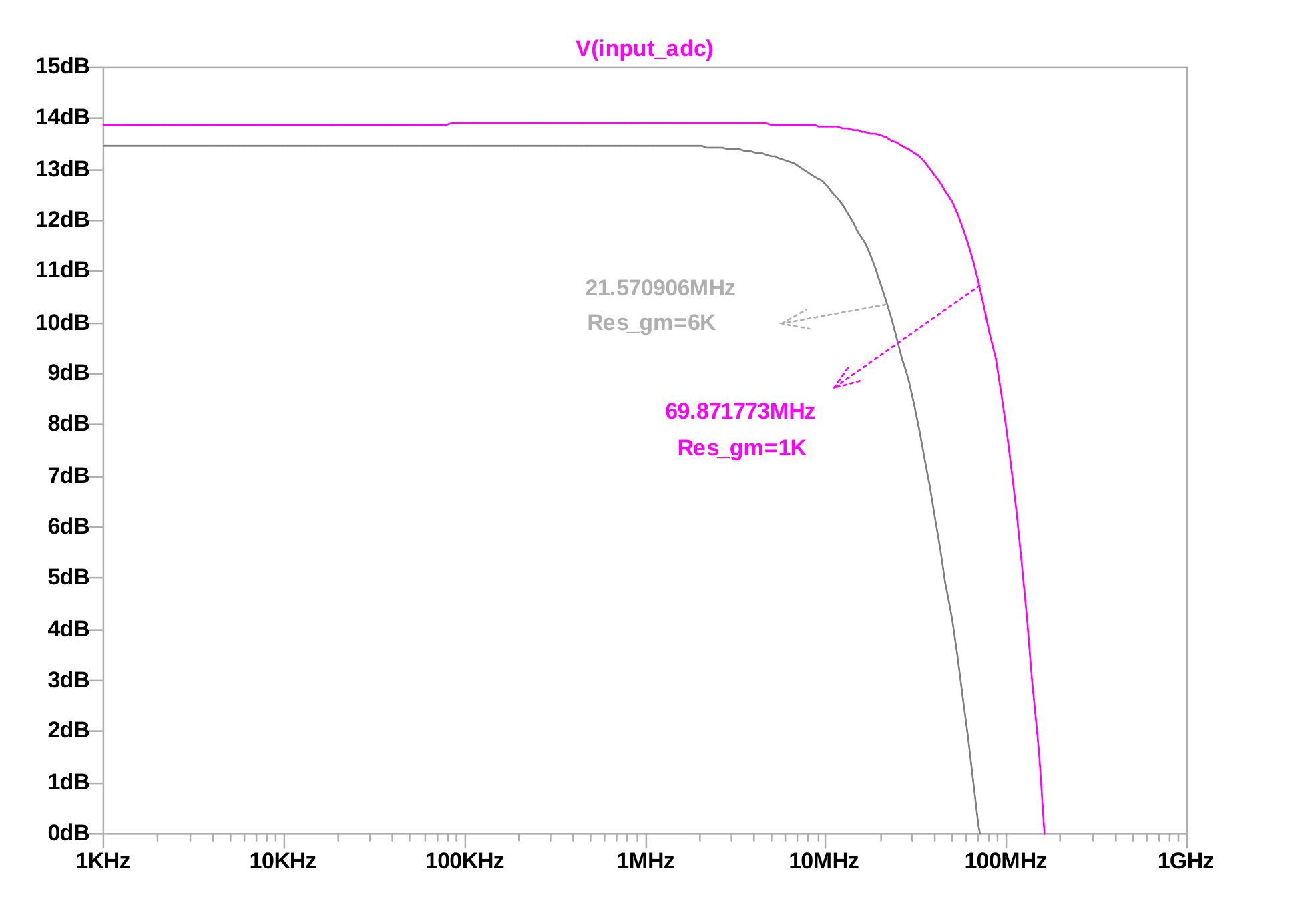}
    \caption{}
    \end{subfigure}
    \begin{subfigure}{.44\textwidth}
    \includegraphics[width=\linewidth]{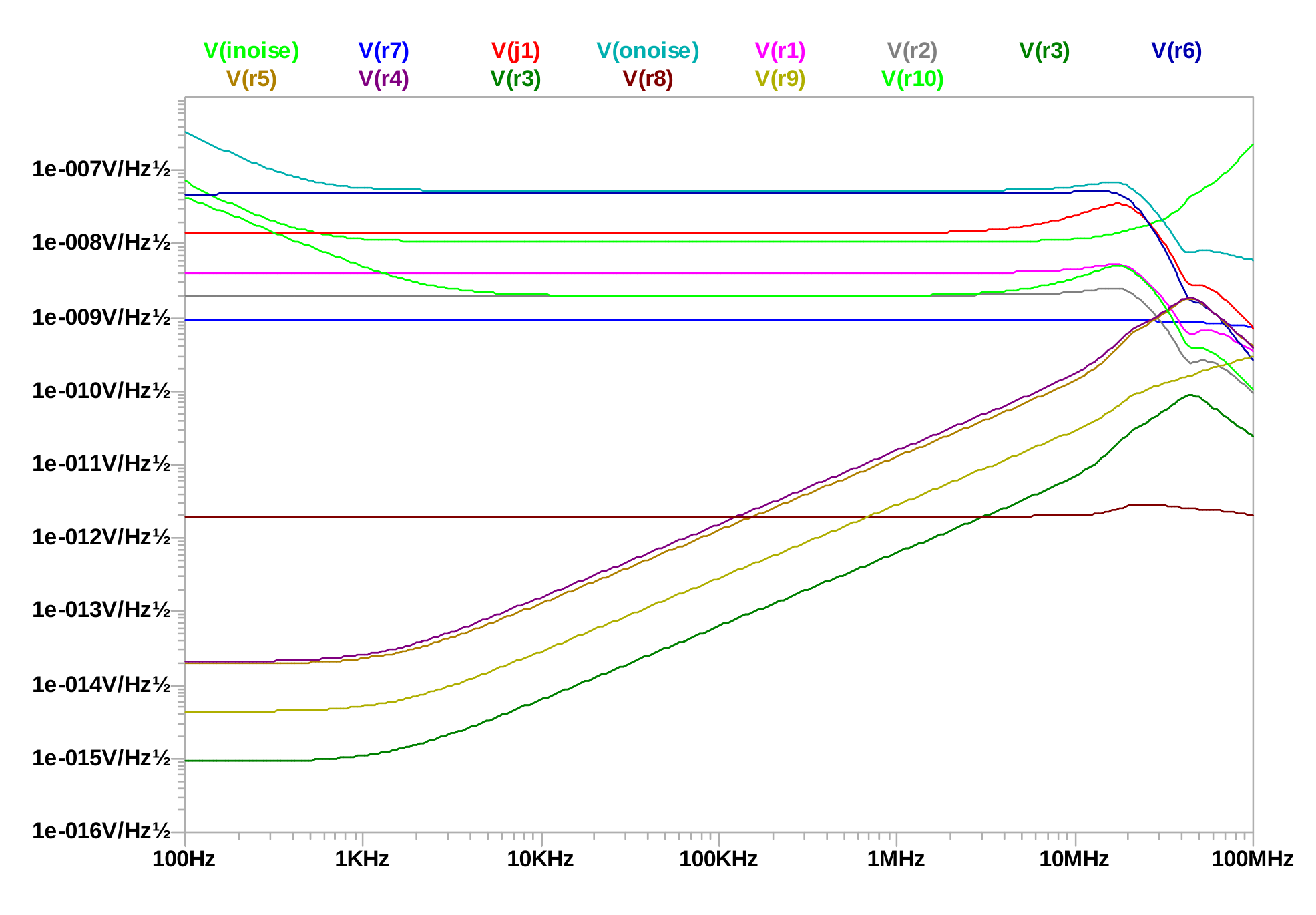}
    \caption{}
    \end{subfigure}
    \begin{subfigure}{.44\textwidth}
    \includegraphics[width=\linewidth]{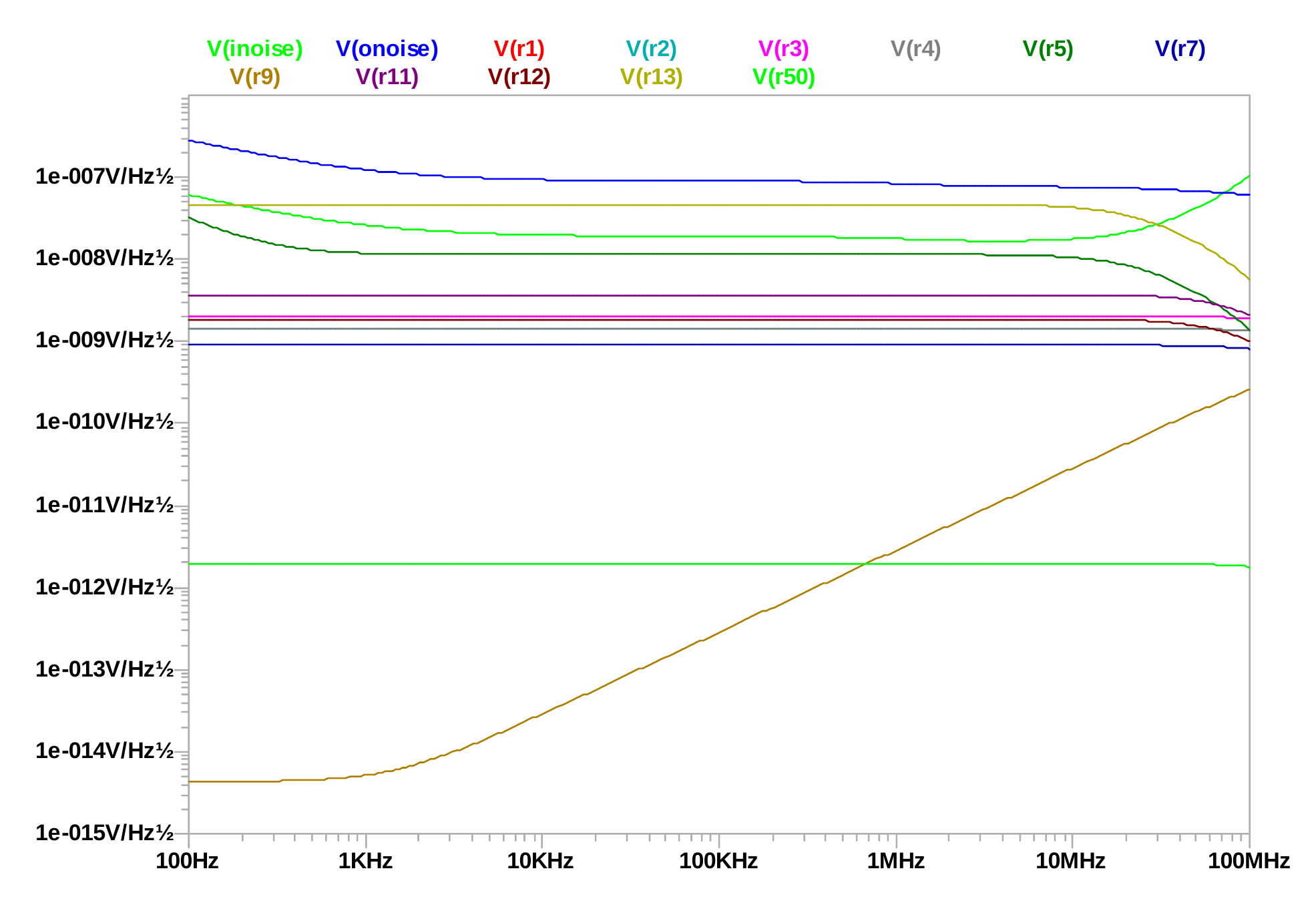}
    \caption{}
    \end{subfigure}
    \caption{Results of simulation in LTspice for the two preamplifiers, C1 (left column) and C2 (right column). (a) and (b) Output from different stages of the preamplifier chains for a 2.5 MHz pulsed input signal. (c) and (d) Predicted bandwidth for the two configurations for two different CCD on-chip amplifier transconductance values (g$_m$=166 $\mu$S and 1 mS). (e) and (f) Predicted noise performance of C1 and C2. Contributions from different components to the total noise are shown in different colors.}
    \label{sims}
\end{figure}

The simulation results for these two configurations are shown in Fig. \ref{sims}.      
The results shown in the left column are obtained for C1 whereas the results in the right column stand for C2. The simulations were done for two different resistor values at the input of the first stage $-$ 6 k$\Omega$ and 1 k$\Omega$, representing the CCD on-chip amplifier g$_m$ of 166 $\mu$S and 1 mS respectively. The g$_m$ of the CCID85 devices is expected to be between these limits.
Fig. \ref{sims}a shows the signal output from the source follower, LSK189 (shown in Cyan), inverting and non-inverting opamps, ADA4896 (shown in red and green respectively) for a pulsed input signal of 500 mV amplitude and 2.5 MHz frequency.  
The overall bandwidth of C1, shown in (c), is mostly independent of this resistor. The 3 dB bandwidth is around 21 MHz for 6 k$\Omega$ and around 28 MHz for 1 k$\Omega$ resistance. For C2, the 1 k$\Omega$ resistor yields a significantly higher bandwidth ($\sim$70 MHz), due to the relatively lower RC time constant of the circuit ($\sim$21 MHz), as shown in panel (d). The effect of lower RC time constant can also be seen in panel (b), where the outputs of the first stage, ADA4817 and differential driver, AD8138 are clearly seen to be slowed down for the higher resistance value (therefore higher RC constant). Panels (e) and (f) demonstrate the noise performance of C1 and C2. For C1, we expect around 10 nV / $\sqrt{Hz}$ noise density at the input whereas for C2, the noise is slightly higher $\sim$20 nV / $\sqrt{Hz}$. These values include the CCD on-chip amplifier noise. Both C1 and C2, therefore, fulfill the low noise requirement of the readout module. 

The preamplifier board accommodates both C1 and C2 preamplifier configurations. In our current design of the board, the P-JFET output node of the device can be read out by one of these preamplifier configurations (C1 or C2) at a time. In this paper, we only discuss the results for C1 preamplifier readout. 
The preamplifier board also accommodates a readout circuit for the SiSeRO current readout and is compatible with both P-JFET and SISERO amplifier outputs such that only one output node (P-JFET or SiSeRO), out of total four output amplifiers, can be read at a time either by P-JFET preamplifier configuration (C1 or C2) or SiSeRO readout chain respectively.    

\subsubsection{Archon controller}

\begin{figure}
    \centering
    \begin{subfigure}{.41\textwidth}
    \includegraphics[width=\linewidth]{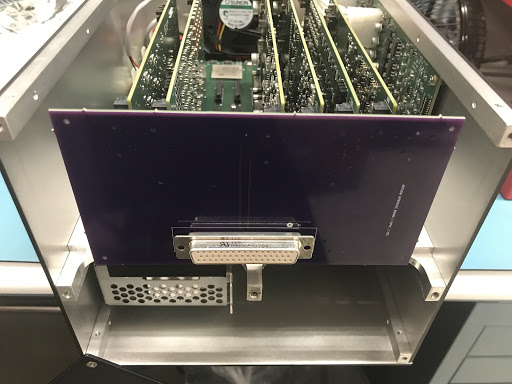}
    \caption{}
    \end{subfigure}
    \begin{subfigure}{.41\textwidth}
    \includegraphics[width=\linewidth]{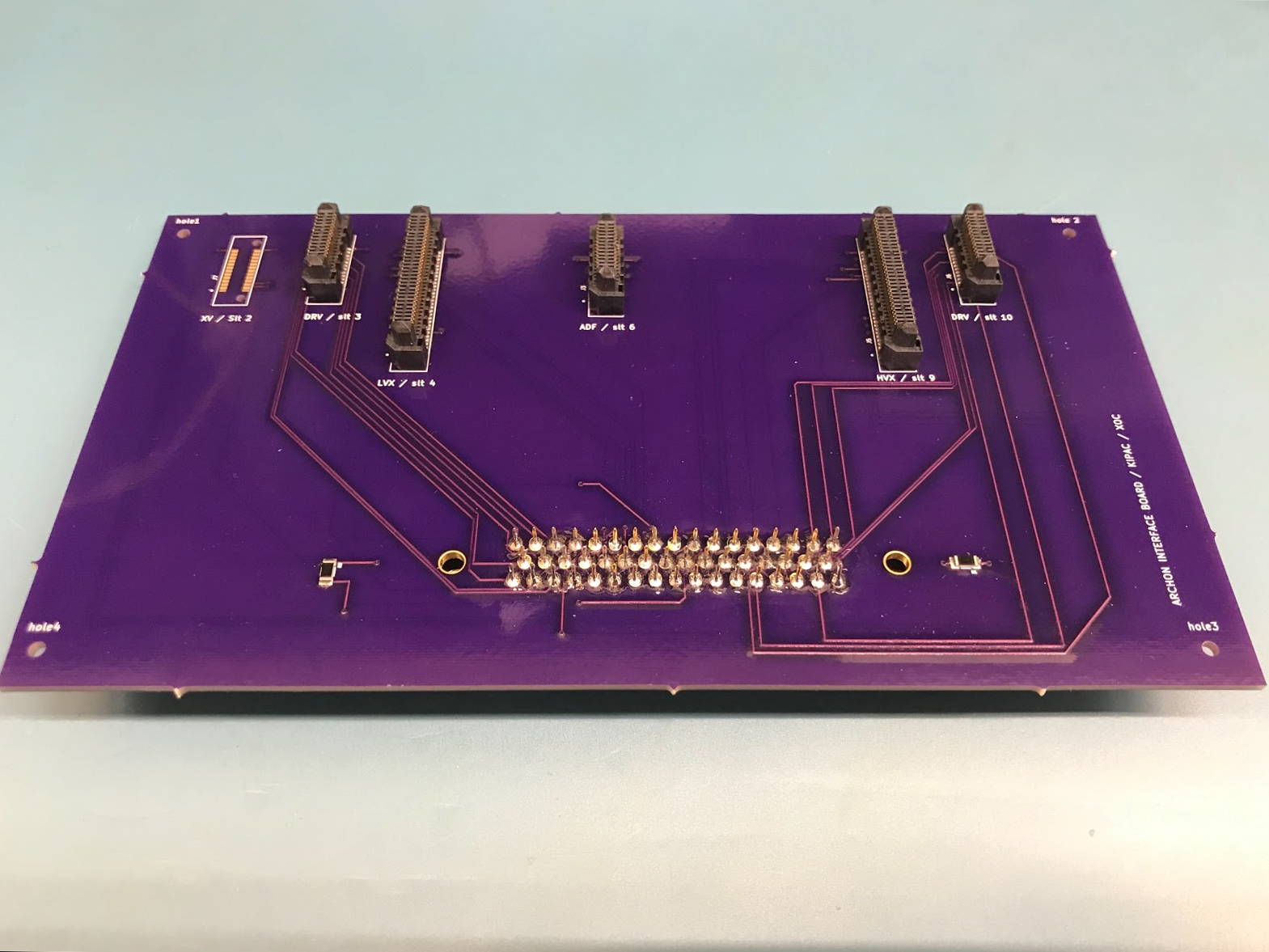}
    \caption{}
    \end{subfigure}
    \caption{(a) The Archon controller and (b) a custom made interface board used to route bias and clocks from the Archon to CCD, and preamplifier output signals to Archon.}
    \label{archon}
\end{figure}

An Archon controller \citep{archon14}, procured from Semiconductor Technology Associates, Inc (STA\footnote{http://www.sta-inc.net/archon/}), provides the next stage of the readout module. This is an FPGA-based modular high performance CCD controller, which provides the necessary bias and clock signals to the CCD, digitizes the CCD output and extracts the signal charge information from the digitized CCD waveform. We also utilize some of the biases for the amplifiers in the preamplifier chains. 
Fig. \ref{archon}a shows the Archon controller along with an interface board. 
The Archon provides a total of 12 slots for ADC, clock driver, bias, heater, or other custom modules, with up to 4 ADC modules for 16 total CCD outputs. In the Archon controller, we use 2 ADC Modules (8 differential ADCs), 2 Clock Modules (16 clock drivers), 1 High Voltage Bias Module (HV) and 1 Low Voltage Bias module (LV). CCD outputs are digitized by 16-bit 100 MHz ADCs. CCD clocks are generated by 14-bit 100 MHz DACs. The LV module provides a total of 30 biases (-14 V to 14 V). The HV Bias Module also provides 30 total biases in 0$-$31 V range.
The Archon receives configuration information about the CCD, e.g. bias signals, clock signals, clocking sequence, sampling of digitized waveform to generate an image, from a host PC and then returns the status and image data to the host PC via a gigabit Ethernet connection. The clocking sequence to transfer the charge from the imaging section of the CCD to the serial register and from there to the output amplifier is done using a timing script which also defines the readout speed of the device.  

The preamplifier board described previously is connected to the Archon through a custom interface board, as shown in Fig. \ref{archon}b. The interface board routes signals from the preamplifier board to the internal Archon module connectors and routes clock and biases from Archon to the CCD through the preamplifier board. 


\section{Characterization of CCID85 detectors}
\label{results}

\begin{figure}
    \centering
    \includegraphics[width=.5\linewidth]{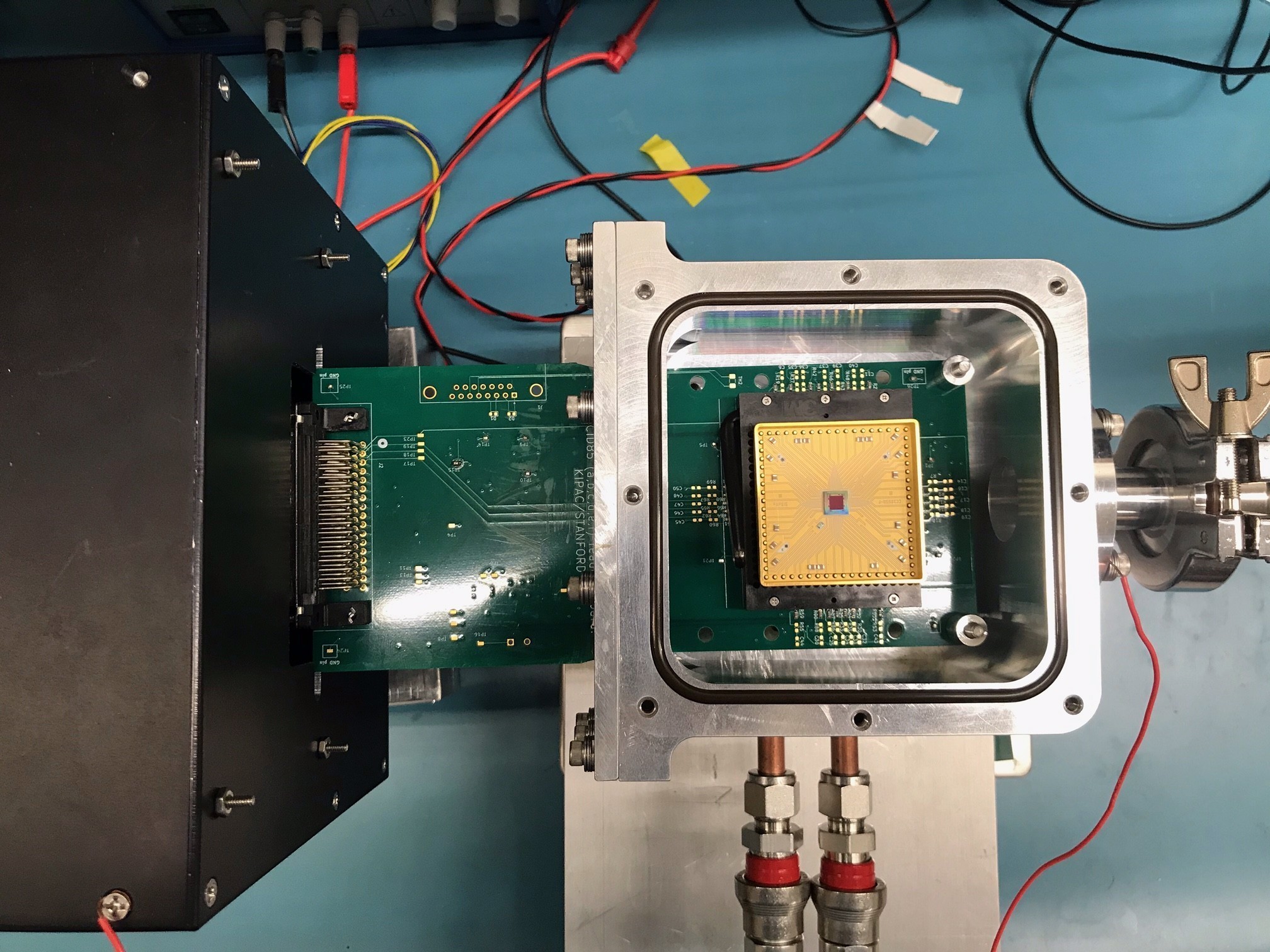}
    \caption{Final experimental set up with a CCID85 device mounted inside the chamber.}
    \label{setup}
\end{figure}

The final experiment setup is shown Fig. \ref{setup}. We mounted a CCID85 device on the PGA socket inside the vacuum chamber to characterize the device and other components of the test stand (e.g. cooling, readout module etc.) 
For the testing, the imaging and serial gates were merged in the preamplifier board such that half of the device (256 $\times$ 512 array) was read out through a P-JFET amplifier output.  
The P-JFET output was processed through the C1 preamplier (see subsection \ref{preamp}) chain and fed to the Archon differential ADCs. We set the serial clock transfer rate (therefore the readout speed) at 2 MHz and imaging clock transfer rate at 350 kHz. 
The substrate bias was set at -10 V for the device.
In the next subsections, we discuss the analysis steps and discuss the spectral resolution and noise performance of the device.      

\subsection{Experimental data and analysis procedure:}

\begin{figure}
    \centering
    \begin{subfigure}{.48\textwidth}
    \includegraphics[width=\linewidth]{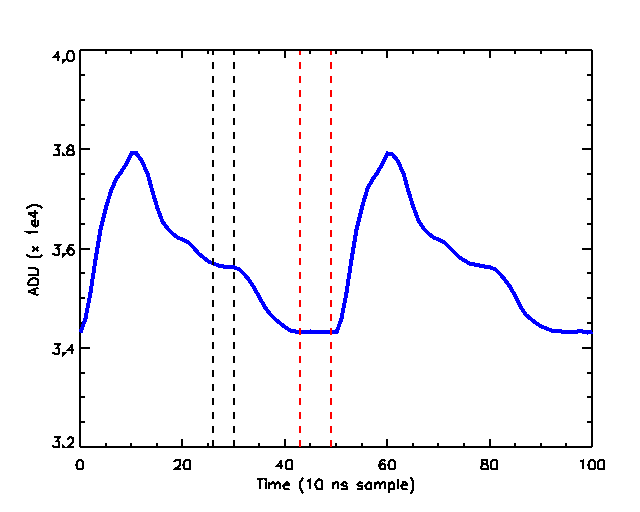}
    \caption{}
    \end{subfigure}
    \begin{subfigure}{.34\textwidth}
\includegraphics[width=\linewidth]{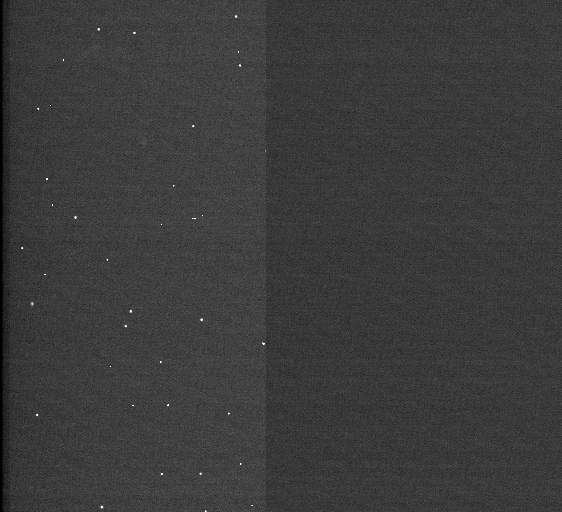}
\caption{}
\end{subfigure}
    \caption{(a) CCD output video waveform obtained from the Archon controller. The regions within the black and red dashed lines represent the baseline and signal levels respectively. The difference of the two levels is proportional to the source signal. (b) An X-ray image obtained from the CCID85 device.}
    \label{waveform}
\end{figure}

The device was cooled down to 233 K (-40$^\circ$C). To test the noise performance of the system, we collected dark frames at four different temperatures, from 263 K to 233 K, at 10 K intervals for multiple frame integration times (0 ms to 5 seconds). An $^{55}\mathrm{Fe}$ radioisotope was used to evaluate the spectral resolution of the device at 5.9 (Mn k$_\alpha$) and 6.4 keV (Mn K$_\beta$) X-ray photons. Fig. \ref{waveform}a shows the CCD output video waveform sampled at every 10 ns as obtained from the Archon controller. 
Every pixel of the CCD generates such a video waveform at the output. 
The change from the baseline (denoted by the black lines) to the signal level (denoted by the red lines), after the clocking transfers the charge from the serial register to the gate of the first stage, is directly proportional to the amount of charge transferred.
The signal amplitude is extracted by taking the difference between the signal and baseline levels (Correlated Double Sampling process or CDS). The 2D images are formed from the computed pixel charges.  
We apply bias and dark frame correction on the raw images to generate cleaned X-ray images. Fig. \ref{waveform}b shows an example of 562 $\times$ 512 array image. The half of the device read out through the output amplifier shows X-ray events. The pre-scan (10 $\times$ 512) and over-scan regions (50 $\times$ 512) are located at the left and rightmost parts of the image respectively. 

\subsection{Read noise}
\begin{figure}
    \centering
    \includegraphics[width=.5\linewidth]{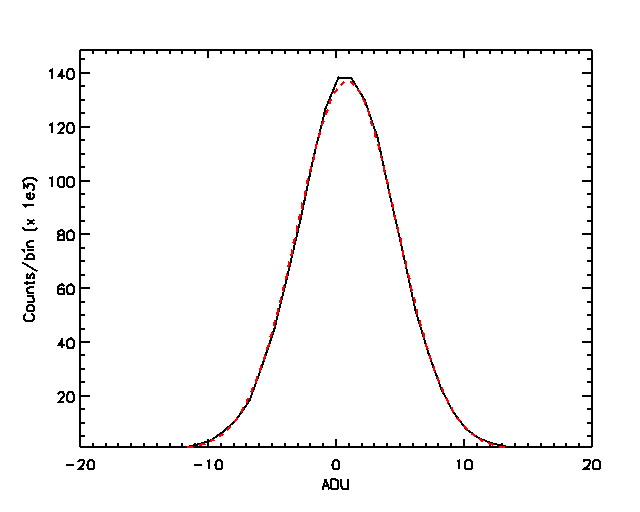}
    \caption{Distribution of charge in ADU from the over-scan region. The red dashed line is a Gaussian fit to the distribution. The read noise is estimated be around 6.95 electrons RMS.}
    \label{readnoise}
\end{figure}

Read noise is estimated from the distribution of charge in the over-scan region. 
An example of such distribution in ADU (Analog to Digital Unit) obtained at 233 K is shown in Fig. \ref{readnoise}.
The distribution is fitted with a Gaussian (shown in red dashed line) to quantify the RMS of the distribution. From the conversion gain of $\sim$2 electrons/ADU for the system, the read noise is estimated to be around 6.95 electrons RMS.  

\begin{figure}
    \centering
    \includegraphics[width=0.8\linewidth]{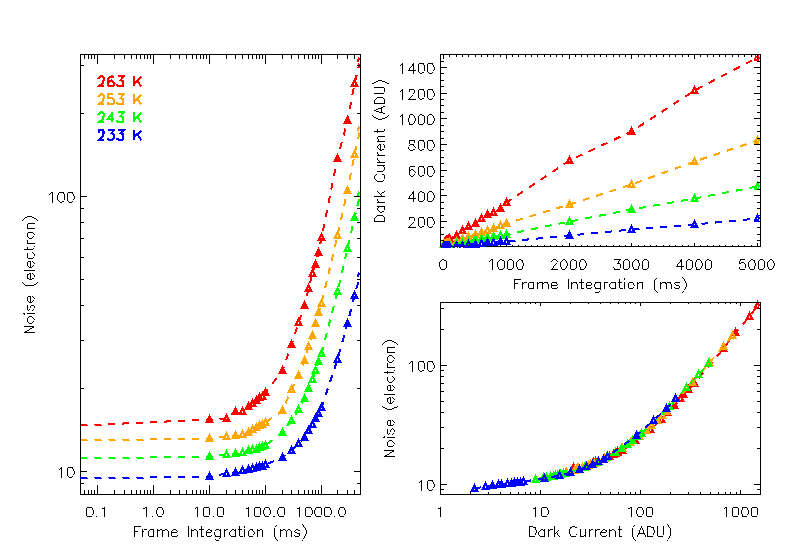}
    \caption{Left: Total system noise as a function of integration time for different detector temperatures. Right-top: dark current as a function of frame integration time. Right-bottom: variation of system noise as a function of dark current.}
    \label{noise}
\end{figure}

The total noise of the system is calculated in the same way by estimating the RMS of charge distribution from the imaging area (active half of the device).  
Fig. \ref{noise} (left) shows the variation of the total noise in electrons as a function of frame integration time and detector temperature. At lower integration times, the total noise is dominated by the read noise of the system, after which the noise is primarily contributed by the leakage current. The total noise at 233 K and 0 ms frame integration is found to be around 9 electrons RMS.
Variation of the mean of dark current as a function of time and temperature is also shown in the figure, along with the variation of the noise with dark current.

\subsection{X-ray spectrum}
\begin{figure}
    \centering
    \begin{subfigure}{.48\textwidth}
    \includegraphics[width=\linewidth]{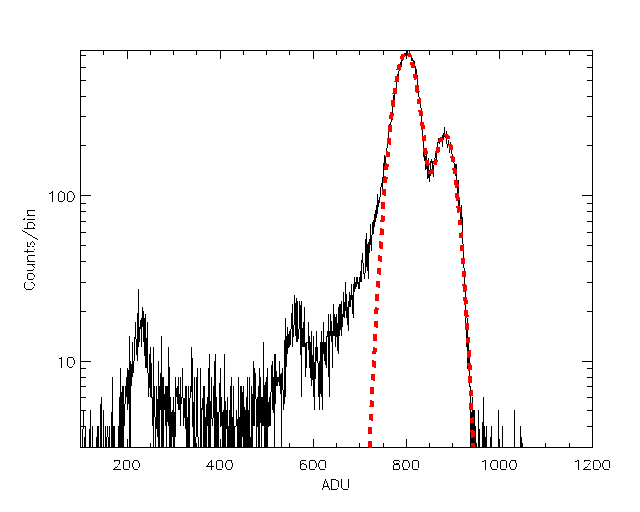}
    \caption{}
    \end{subfigure}
    \begin{subfigure}{.48\textwidth}
    \includegraphics[width=\linewidth]{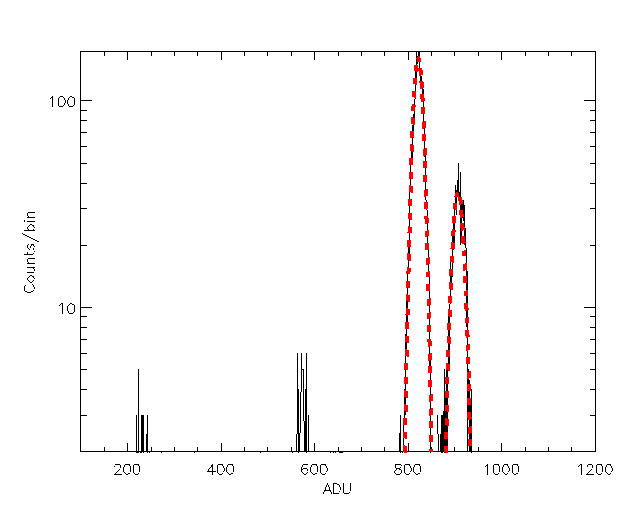}
    \caption{}
    \end{subfigure}
    \caption{A spectrum showing the Mn K$_{\alpha}$ (5.9 keV) and K$_{\beta}$ (6.4 keV) lines from a $^{55}$Fe radioactive source for (a) all pixel events and (b) single-pixel (grade 0) events. Full Width at Half Maximums (FWHM) are estimated to be around 410 eV and 162 eV at 5.9 keV for (a) and (b) respectively.}
    \label{spectrum}
\end{figure}

An X-ray spectrum showing Mn K$_\alpha$ (5.9 keV) and Mn K$_\beta$ (6.4 keV) lines from a $^{55}\mathrm{Fe} $ radioactive source is shown in Fig. \ref{spectrum}, where (a) and (b) stand for all-pixel events and single-pixel events (grade 0 events) respectively. The red dashed lines are Gaussian fits to the spectra.
An event list of 9-pixel islands is generated around each X-ray event. We apply a secondary ADU threshold $\sim$3$\times$read noise to determine the 1-pixel, 2-pixel and multi-pixel events to generate spectra. We calculate a full width at half maximum (FWHM) of 410 eV at 5.9 keV (6.95 \% energy resolution) from the all-pixel event spectra. The energy resolution improves significantly for the single-pixel events, to 162 eV FWHM (2.85 \% energy resolution). Because of the small pixel sizes of the device ($\sim$8 $\mu$m), only a small fraction of the X-ray events are expected to contribute to the single pixel spectrum. We also see an escape peak for 5.9 keV photons and aluminum K$_\alpha$ line (1.5 keV) at the lower end of Fig. \ref{spectrum}a and b.   

\section{Summary and future plans}

We have discussed a new test stand (``tiny-box") developed at Stanford University to characterize X-ray CCDs and develop fast low noise readout electronics for future X-ray astronomy missions. Capability of the test stand was demonstrated by characterizing an X-ray CCD at 2 MHz readout speed and multiple temperatures. Preliminary results from the P-JFET amplifier of the CCD output are very promising. In the future, we plan to carry out detailed characterization of the CCDs at higher readout speeds ($>$ 5 MHz). Another goal of the project is to optimize the readout circuits for SiSeRo amplifier outputs of the CCDs and demonstrate the speed-noise performance of these devices. We have also started to develop an ASIC-based readout system at Stanford University to enable multiple / parallel CCD output readout, the details of which can be found in Herrmann et al., 2020 (in this issue, Paper No.	11454-85).

\acknowledgments 

This work has been supported by APRA $\#$80NSSC19K0499 ``Development of Integrated Readout Electronics for Next Generation X-ray CCDs".


\end{document}